\documentclass[aps,prd,preprint,12pt,superscriptaddress,nofootinbib,a4paper]{revtex4-1}

\usepackage[utf8]{inputenc}
\usepackage{amsmath,amssymb}
\usepackage[separate-uncertainty=true]{siunitx}
\usepackage{graphicx}
\usepackage[usenames,dvipsnames]{xcolor}
\usepackage[caption=false]{subfig}
\usepackage{hyperref}
\usepackage[capitalize]{cleveref}
\usepackage{booktabs}
\usepackage{tabularx}
\usepackage{xspace}
\usepackage{color}
\usepackage[normalem]{ulem}
\usepackage{slashed}
\usepackage{bbold}
\usepackage{wasysym}
\usepackage{graphicx}
\usepackage{mathrsfs} 

\allowdisplaybreaks

\graphicspath{{graphics/}}

\newcommand{\eqdot}{\,.}
\newcommand{\eqcomma}{\,,}
\newcommand{\eqn}{equation}
\newcommand{\lb}{\left(}
\newcommand{\rb}{\right)}

\newcommand{\al}{\alpha}

\newcommand{\Ztwo}{\ensuremath{{\mathbb{Z}_2}}\xspace}
\newcommand{\HiggsBounds}{\texttt{HiggsBounds}\xspace}
\newcommand{\HiggsSignals}{\texttt{HiggsSignals}\xspace}
\newcommand{\HSv}[1]{\texttt{HiggsSignals-#1}}
\newcommand{\HBv}[1]{\texttt{HiggsBounds-#1}}
\newcommand{\HS}{\texttt{HiggsSignals}}
\newcommand{\HB}{\texttt{HiggsBounds}}

\newcommand{\GeV}{{\ensuremath\rm GeV}}
\newcommand{\TeV}{{\ensuremath\rm TeV}}

\newcommand{\pb}{{\ensuremath\rm pb}}
\newcommand{\fb}{{\ensuremath\rm fb}}

\DeclareSIUnit{\pb}{pb}
\DeclareSIUnit{\fb}{fb}
\AtBeginDocument{
\heavyrulewidth=.08em
\lightrulewidth=.05em
\cmidrulewidth=.03em
\belowrulesep=.65ex
\belowbottomsep=0pt
\aboverulesep=.4ex
\abovetopsep=0pt
\cmidrulesep=\doublerulesep
\cmidrulekern=.5em
\defaultaddspace=.5em

\newcolumntype{C}{>{\centering\arraybackslash}X}
\newcolumntype{b}{C}
\newcolumntype{s}{>{\hsize=.6\hsize}C}
\newcolumntype{R}{>{\raggedleft\arraybackslash}X}
}

\begin{document}
\bibliographystyle{hunsrt}
\date{\today}
\rightline{RBI-ThPhys-2022-35}
\title{{\Large Two-Real-Singlet-Model Benchmark Planes}}

\author{Tania Robens}
\email{trobens@irb.hr}
\affiliation{Ruder Boskovic Institute, Bijenicka cesta 54, 10000 Zagreb, Croatia}

\renewcommand{\abstractname}{\texorpdfstring{\vspace{0.5cm}}{} Abstract}

\begin{abstract}
    \vspace{0.5cm}
In this manuscript, I briefly review the Benchmark Planes in the Two-Real-Singlet Model (TRSM), a model that enhances the Standard Model (SM) scalar sector by two real singlets that obey a $\mathbb{Z}_2\,\otimes\,\mathbb{Z}_2'$ symmetry. In this model, all fields acquire a vacuum expectation value, such that the model contains in total 3 CP-even neutral scalars that can interact with each other. All interactions with SM-like particles are inherited from the SM-like doublet via mixing. I remind the readers of the previously proposed benchmark planes, and briefly discuss possible production at future Higgs factories, as well as regions in a more generic scan of the model. For these, I also discuss the use of the W-boson mass as a precision observable to determine allowed/ excluded regions in the models parameter space. This work is an extension of a whitepaper submitted to the Snowmass process.
  
\end{abstract}

\maketitle


\section{Introduction and Model}

After the discovery of a scalar boson that complies very well with the predictions for the Standard Model (SM) Higgs sector (see e.g. \cite{ATLAS:2022vkf,CMS:2022dwd} for recent experimental summaries), particle physics has entered an exciting era. One crucial question is whether the scalar we are observing indeed corresponds to the Higgs boson predicted by the SM, or whether it is part of an extended scalar sector, introducing additional scalar states. For many years, many models have been suggested that extend the SM scalar sector by additional electroweak singlets, doublets, or other multiplets.

From a bottom up approach, the easiest extension of the SM scalar sector is the extension by an additional gauge singlet, where further symmetries can be imposed in order to reduce the number of free parameters in the model. Such extensions have been e.g. discussed in \cite{Pruna:2013bma,Robens:2015gla,Robens:2016xkb}, with a more recent update on the allowed parameter space in \cite{Robens:2022oue}. Such models can also allow for a strong first-order electroweak phase transitions, see e.g. discussion in \cite{Papaefstathiou:2022oyi,Robens:2022zgk} and references therein. Discussion of such models without an additional symmetry can e.g. be found in \cite{Lewis:2017dme,Chen:2017qcz}

The simple singlet extensions only allow for one additional scalar. Experimental searches, however, now start to investigate so called non-symmetric production modes of the form

\begin{\eqn*}
p\,p\,\rightarrow\,h_a\,\rightarrow\,h_b\,h_c,
\end{\eqn*}
where $h_{a,b,c}$ here denote scalar states with different masses; first results for such searches have been presented in \cite{CMS:2022suh,CMS-PAS-HIG-21-011}. To allow for such final states, at least one additional scalar needs to be among the particle content of the considered model. Although many new physics extensions allow for such scenarios, again the most straightforward realization is a model where one additional scalar field is added that transforms as a singlet under the SM gauge group. This is the model that this work focusses on.

The model discussed here has been proposed in \cite{Robens:2019kga}, and I refer the reader to that reference for a detailed discussion of model setup and constraints. I just briefly repeat the generic features for completeness. The work presented here is an extension of a Snowmass white paper \cite{Robens:2022lkn}. Similar models have e.g. been discussed in ~\cite{Barger:2008jx,AlexanderNunneley:2010nw,Coimbra:2013qq,Ahriche:2013vqa,Costa:2014qga,Costa:2015llh,Ferreira:2016tcu,Chang:2016lfq,Muhlleitner:2017dkd,Dawson:2017jja,aali:2020tgr,Ghosh:2022cca,Adhikari:2022yaa}.

The potential in the scalar sector is given by
\begin{equation}
    \begin{aligned}
        V\lb \Phi,\,S,\,X\rb & = \mu_{\Phi}^2 \Phi^\dagger \Phi + \lambda_{\Phi} {(\Phi^\dagger\Phi)}^2
        + \mu_{S}^2 S^2 + \lambda_S S^4
        + \mu_{X}^2 X^2 + \lambda_X X^4                                              \\
          & \quad+ \lambda_{\Phi S} \Phi^\dagger \Phi S^2
        + \lambda_{\Phi X} \Phi^\dagger \Phi X^2
        + \lambda_{SX} S^2 X^2\eqdot
    \end{aligned}\label{eq:TRSMpot}
\end{equation}
Here, $\Phi$ denotes the SM-like doublet, while $X,\,S$ are two additional real scalar fields. The model obeys an additional $\Ztwo\,\otimes\,\Ztwo'$ symmetry
$        \Ztwo: \, S\to -S\eqcomma
        \Ztwo ': \, X\to -X,$ while all other fields transform evenly under the respective $\Ztwo$ symmetry. All three scalars acquire a vacuum expectation value (vev) and therefore mix. This leads to three physical states with all possible scalar-scalar interactions.

Among the important constraints are e.g. the Higgs signal strength measurements by the LHC experiments, perturbative unitarity as well as the requirement for the potential to be bounded from below, and current collider searches. Results have been obtained using the \texttt{ScannerS}~\cite{Coimbra:2013qq,Ferreira:2014dya,Costa:2015llh,Muhlleitner:2016mzt,Muhlleitner:2020wwk} framework. Experimental results from past and current collider experiments have been implemented using the publicly available tools \HiggsBounds \cite{Bechtle:2008jh,Bechtle:2011sb,Bechtle:2013gu,Bechtle:2013wla, Bechtle:2015pma,Bechtle:2020pkv} and \HiggsSignals \cite{Stal:2013hwa, Bechtle:2013xfa,Bechtle:2014ewa,Bechtle:2020uwn}.

In the following, we will use the convention that
\begin{\eqn}\label{eq:hier}
M_1\,\leq\,M_2\,\leq\,M_3
\end{\eqn}
and denote the corresponding physical mass eigenstates by $h_i$.
Gauge and mass eigenstates are related via a mixing matrix. The model contains in total 9 free parameters, out of which 2 are fixed by the observation of a scalar particle with the mass of 125 \GeV~ as well as electroweak precision observables. Apart from the masses, also the vaccum expectation values (vevs) and mixing angles serve as input parameters. Interactions with SM particles are then inherited from the scalar excitation of the doublet via rescaling factors $\kappa_i$, such that $g_i^{h_i A B}\,=\,\kappa_i\,g_i^{h_i A B,\text{SM}}$ for any $h_i A B$ coupling, where $A,\,B$ denote SM particles. Orthogonality of the mixing matrix implies $\sum_i \kappa_i^2\,=\,1$. Furthermore, signal strength measurements require $|\kappa_{125}|\gtrsim\,0.96$ \cite{Robens:2019kga}\footnote{Note that the Run 2 combinations of ATLAS \cite{ATLAS:2022vkf} and CMS \cite{CMS:2022dwd} separately lead to $|\kappa_{125}|\,\gtrsim\,0.96$  and $|\kappa_{125}|\,\gtrsim\,0.94$, respectively. All benchmark planes in \cite{Robens:2019kga} fulfill these requirements.} for the SM-like scalar $h_{125}$, which can be $h_1,\,h_2$ or $h_3$ depending on the specific parameter choice.

For a certain production process (e.g.~gluon gluon fusion) the cross
section, $\sigma$, for $h_a$ with mass $M_a$ can be obtained from the
corresponding SM Higgs production cross section, $\sigma_\text{SM}$, by
simply rescaling
\begin{equation}
    \sigma(M_a) = \kappa_a^2 \cdot \sigma_\text{SM} ( M_a)\eqdot\label{eq:cxnscaling}
\end{equation}
Since $\kappa_a$ rescales all Higgs couplings to SM particles,
\cref{eq:cxnscaling} is exact up to genuine electroweak corrections involving Higgs
self-interactions, and in particular holds to all orders in QC\@D.

The scaling factor $\kappa_a$ plays the same role in universally rescaling the partial widths of
$h_a$ decays into SM particles, leading to 
\begin{equation}
    \Gamma(h_a\to\text{SM}; M_a) = \kappa_a^2 \cdot \Gamma_\text{tot}(h_\text{SM}; M_a),\label{eq:widthscaling}
\end{equation}
where $\Gamma( h_a \to \text{SM}; M_a)$ denotes the sum of all partial widths of $h_a$ into SM particle final states. In addition, the branching ratios (BRs) of $h_a$ decays to other scalar bosons, $h_a \to h_b h_c$, are given by:
\begin{equation}
    \text{BR}(h_a\to h_b h_c) = \frac{\Gamma_{a\to bc}}{\kappa_a^2~\Gamma_\text{tot}(h_\text{SM}) + \sum_{xy} \Gamma_{a\to xy}}\eqdot
\end{equation}
where the denominator now denotes the total width of the scalar $h_a$. In the absence of BSM decay modes --- which
is always the case for the lightest Higgs bosons $h_1$ --- $h_a$ has BRs
identical to a SM-like Higgs boson of the same mass.
\section{Benchmark planes}
In \cite{Robens:2019kga}, several benchmark planes (BPs) were proposed which were meant to capture mainly features that by the time of that publication were not yet adressed by searches at the LHC:
\begin{itemize}
\item{}asymmetric production and decay, in the form of
\begin{\eqn*}
p\,p\,\rightarrow\,h_3\,\rightarrow\,h_1\,h_2,
\end{\eqn*}
where, depending on the kinematics, $h_2\,\rightarrow\,h_1\,h_1$ decays are also possible;
\item{}symmetric decays in the form of
\begin{\eqn*}
p\,p\,\rightarrow\,h_i\,\rightarrow\,h_j\,h_j,
\end{\eqn*}
where none of the scalars corresponds to the 125 \GeV~ resonance. Note that this in principle allows for further decays $h_j\,\rightarrow\,h_k\,h_k$, again depending on the specific benchmark plane kinematics.
\end{itemize}
We list the definition of these benchmark planes in tables \ref{tab:benchmarkoverview} and \ref{tab:BPparams}, respectively.

\begin{table} 
    \centering
    \begin{tabularx}{\textwidth}{sssb}
        \toprule
        benchmark scenario & $h_{125}$ candidate & target signature      & possible successive decays                          \\
        \midrule
        \textbf{BP1}       & $h_3$               & $h_{125} \to h_1 h_2$ & $h_2 \to h_1 h_1$ if $M_2 > 2 M_1$                  \\
        \textbf{BP2}       & $h_2$               & $h_3 \to h_1 h_{125}$ & -                                                   \\
        \textbf{BP3}       & $h_1$               & $h_3 \to h_{125} h_2$ & $h_2 \to h_{125}h_{125}$ if $M_2 > \SI{250}{\GeV}$  \\
        \textbf{BP4}       & $h_3$               & $h_2 \to h_1 h_1$     & -                                                   \\
        \textbf{BP5}       & $h_2$               & $h_3 \to h_1 h_1$     & -                                                   \\
        \textbf{BP6}       & $h_1$               & $h_3 \to h_2 h_2$     & $h_2 \to h_{125}h_{125}$ if  $M_2 > \SI{250}{\GeV}$ \\
        \bottomrule
    \end{tabularx}
    \caption{Overview of the benchmark scenarios: The second column denotes the
    Higgs mass eigenstate that we identify with the observed Higgs boson,
    $h_{125}$, the third column names the targeted decay mode of the resonantly
    produced Higgs state, and the fourth column lists possible relevant
    successive decays of the resulting Higgs states.}\label{tab:benchmarkoverview}
\end{table}
\begin{table} 
    \centering
    \begin{tabularx}{\textwidth}{CRRRRRR}
        \toprule
        Parameter           & \multicolumn{6}{c }{Benchmark scenario}                                                                             \\
                            & \textbf{BP1}                            & \textbf{BP2} & \textbf{BP3} & \textbf{BP4} & \textbf{BP5} & \textbf{BP6}  \\
        \midrule
        $M_1~[\SI{}{\GeV}]$ & $[1, 62]$                               & $[1,124]$    & $125.09$     & $[1, 62]$    & $[1, 124]$   & $125.09$      \\
        $M_2~[\SI{}{\GeV}]$ & $[1, 124]$                              & $125.09$     & $[126, 500]$ & $[1,124]$    & $125.09$     & $[126, 500]$  \\
        $M_3~[\SI{}{\GeV}]$ & $125.09$                                & $[126,500]$  & $[255, 650]$ & $125.09$     & $[126, 500]$ & $[255, 1000]$ \\
        $\theta_{hs}$       & $1.435$                                 & $1.352$      & $-0.129$     & $-1.284$     & $-1.498$     & $0.207$       \\
        $\theta_{hx}$       & $-0.908$                                & $1.175$      & $0.226$      & $1.309$      & $0.251$      & $0.146$       \\
        $\theta_{sx}$       & $ -1.456$                               & $-0.407$     & $-0.899$     & $-1.519$     & $0.271$      & $0.782$       \\
        $v_s~[\SI{}{\GeV}]$ & $630$                                   & $120$        & $140$        & $990$        & $50$         & $220$         \\
        $v_x~[\SI{}{\GeV}]$ & $700$                                   & $890$        & $100$        & $310$        & $720$        & $150$         \\
        \midrule
        $\kappa_1$          & $0.083$                                 & $0.084$      & $0.966$      & $0.073$      & $0.070$      & $0.968$       \\
        $\kappa_2$          & $0.007$                                 & $0.976$      & $0.094$      & $0.223$      & $-0.966$     & $0.045$       \\
        $\kappa_3$          & $-0.997$                                & $-0.203$     & $0.239$      & $0.972$      & $-0.250$     & $0.246$       \\
        \bottomrule
    \end{tabularx}
    \caption{Input parameter values and coupling scale factors, $\kappa_a$
    ($a=1,2,3$), for the six defined benchmark scenarios. The doublet vev is set
    to $v=\SI{246}{\GeV}$ for all scenarios.}
    \label{tab:BPparams}
\end{table}

For this work, I rescanned all benchmark planes with the newest \HiggsBounds and \HiggsSignals versions: \HBv{5.10.2} and \HSv{2.6.2}. For nearly all parameter points, these new versions did not introduce additional constraints on the parameter space, and I therefore show the benchmark planes from the original publication. One exception is BP5 which has a slightly more constrained parameter space taking additional searches into account. I also comment on a possible recast on this plane and give a list of current experimental searches partially relying on our model. All cross sections which are displayed are for a center-of-mass (COM) energy of 13 \TeV~ and have been derived using rescaled predictions of the NNLO+NNLL production cross sections for a SM-like Higgs of the respective mass as tabulated in \cite{LHCHiggsCrossSectionWorkingGroup:2013rie}, see \cite{Robens:2019kga} for a more detailed discussion.

\subsection{Asymmetric decays}
\begin{center}
\begin{figure} 
\begin{center}
\begin{minipage}{0.45\textwidth}
\begin{center}
\includegraphics[width=\textwidth]{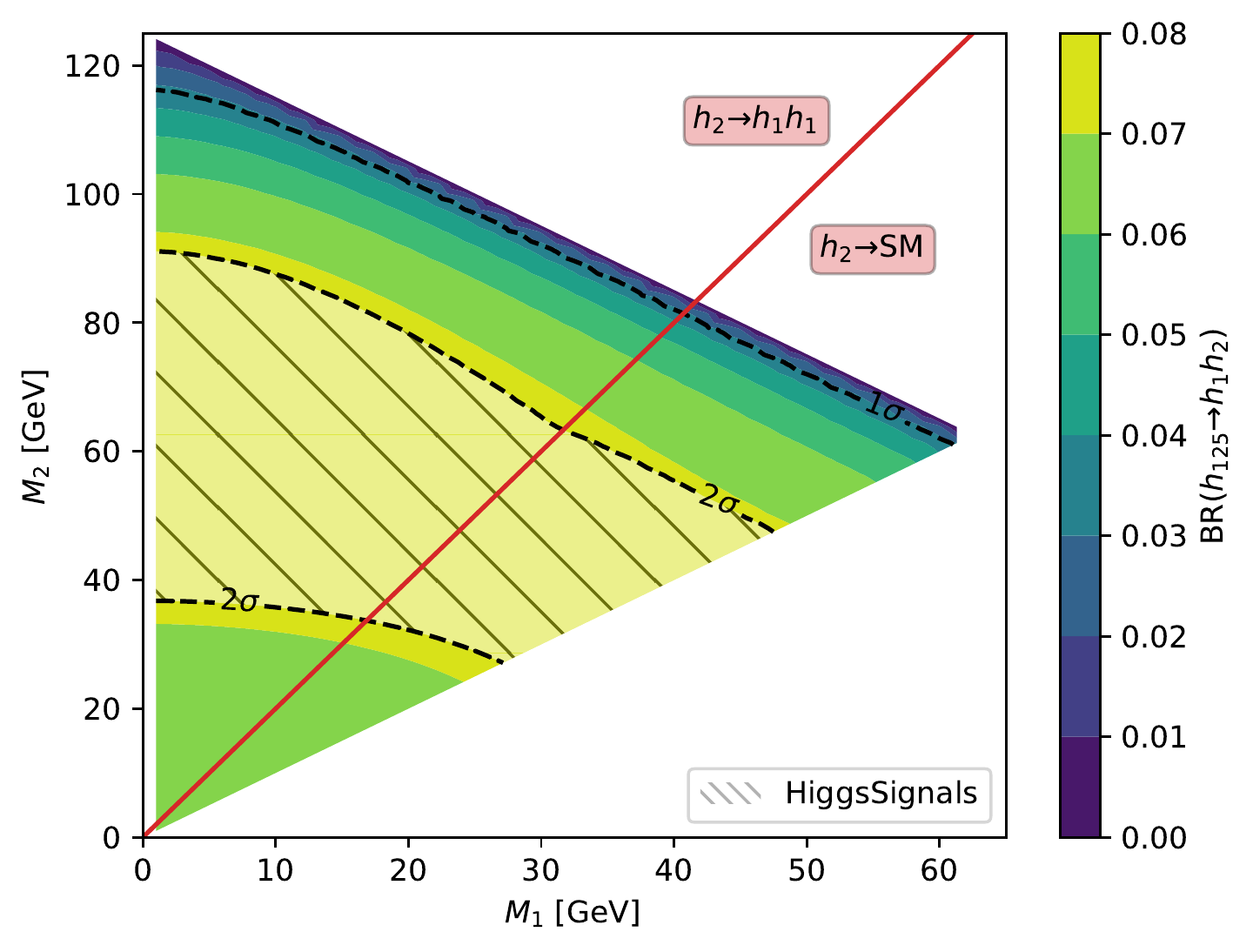}\\
\vspace{-3mm}
{\small Benchmark Plane 1}
\end{center}
\end{minipage}\\ 
\vspace{5mm}
\begin{minipage}{\textwidth}
\begin{center}
\includegraphics[width=0.45\textwidth]{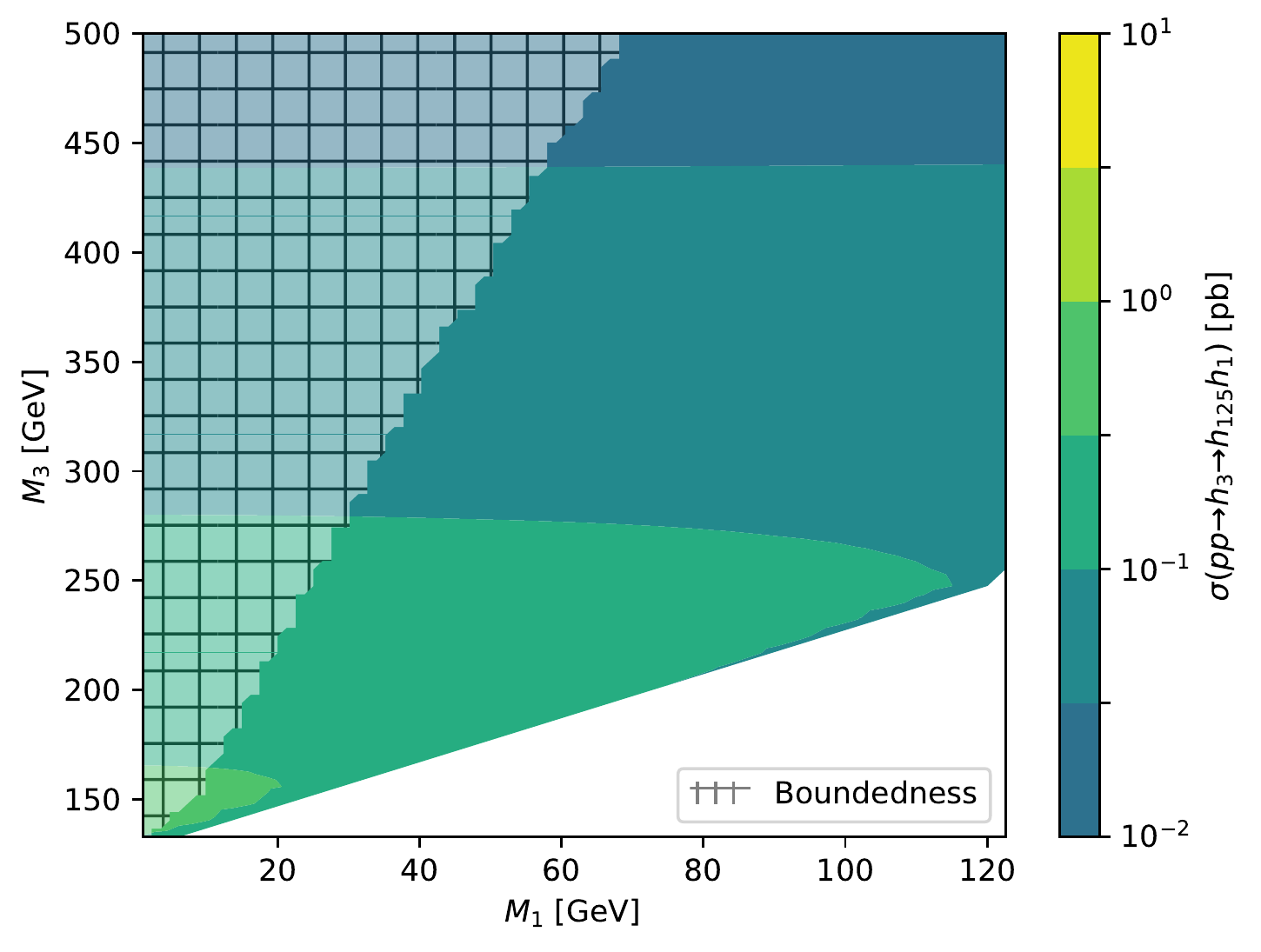}
\includegraphics[width=0.45\textwidth]{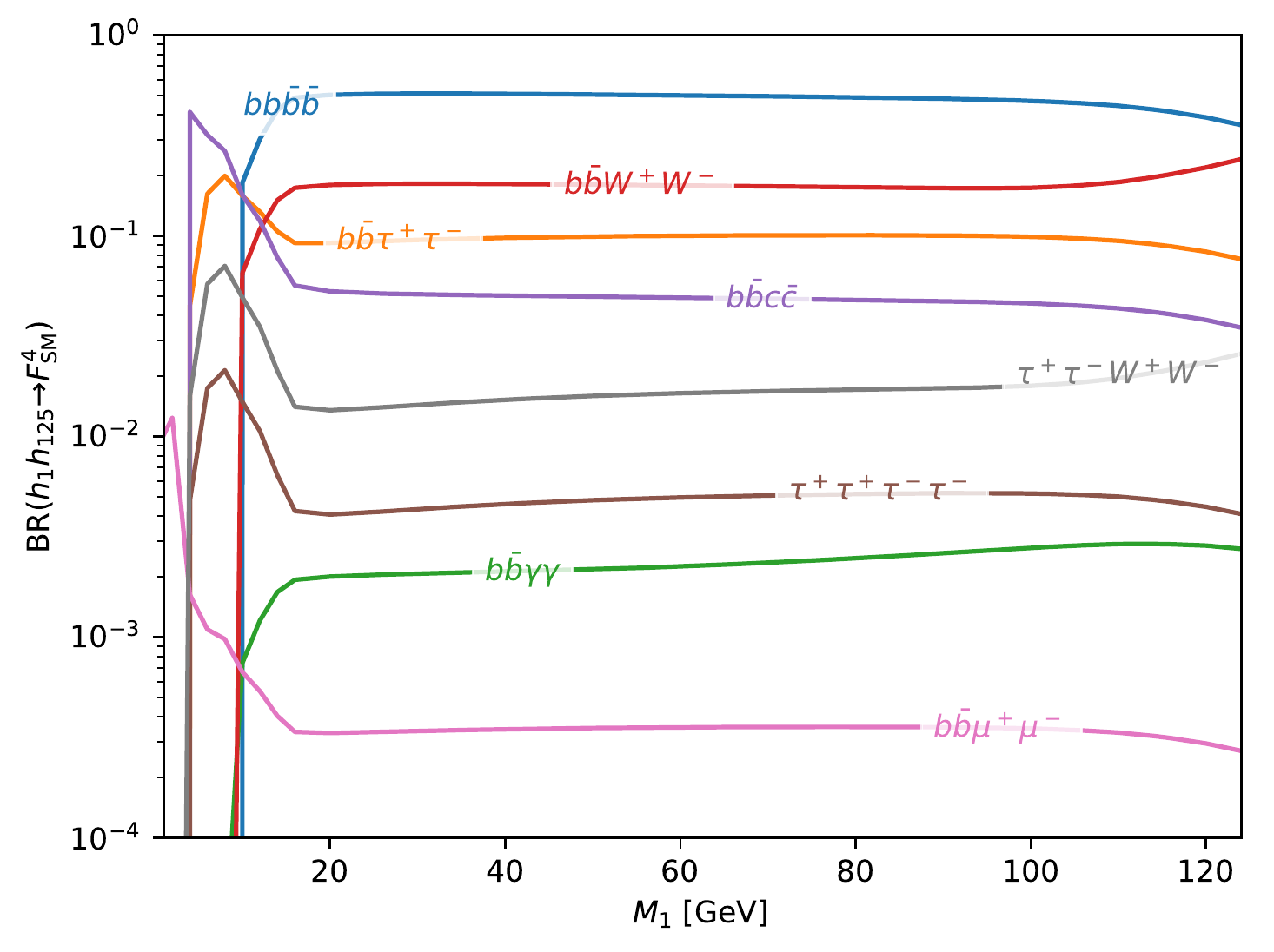}\\
\vspace{-4mm}
{Benchmark Plane 2}
\end{center}
\end{minipage}\\
\vspace{3mm}
\begin{minipage}{\textwidth}
\begin{center}
\includegraphics[width=0.45\textwidth]{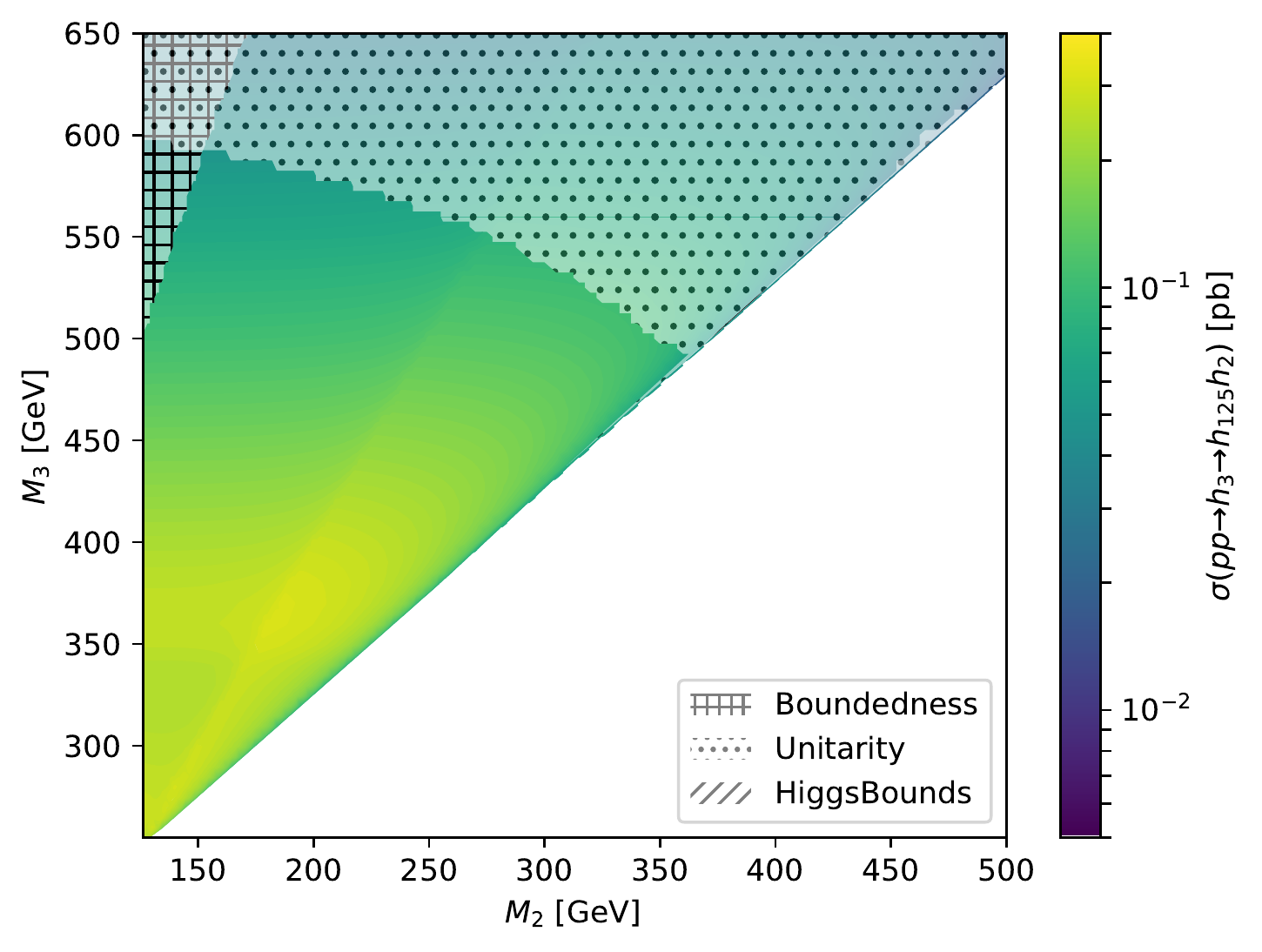}
\includegraphics[width=0.49\textwidth]{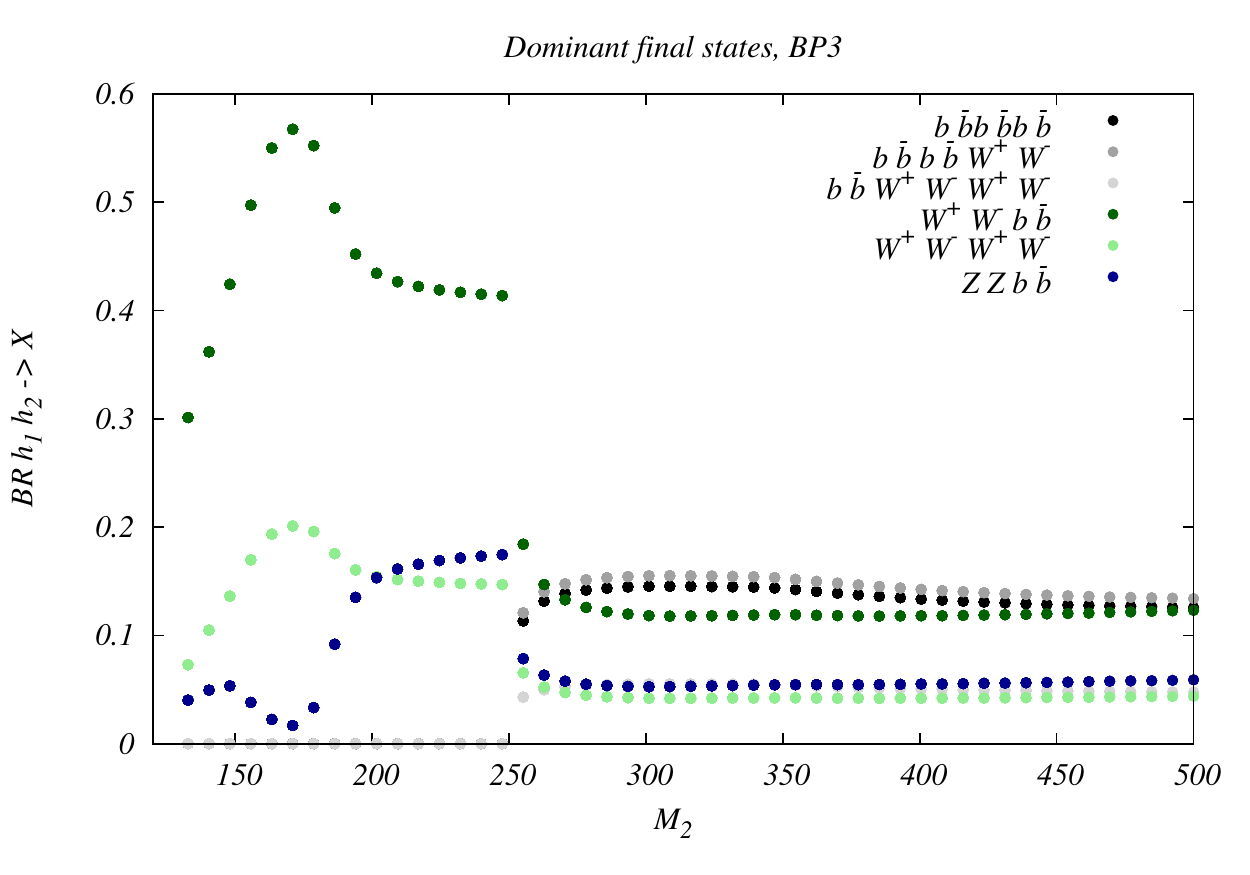}\\
\vspace{-3mm}
{Benchmark Plane 3}
\end{center}
\end{minipage}\\
\end{center}
\caption{\label{fig:asymm} Benchmark planes for asymmetric production and decay, $p\,p\,\rightarrow\,h_{3}\,\rightarrow\,h_1\,h_2$, for various assigments of the 125 \GeV~ resonance. {\sl Top row:} BP1, where $h_3\,\equiv\,h_{125}$. Production cross sections are close to the SM production here, of around $\sim\,4\,\pb$ at 13 \TeV. Shown is the branching ratio to $h_1\,h_2$ in the two-dimensional mass plane. {\sl Middle and bottom rows:} BPs 2 and 3, where $h_{2,1}\,\equiv\,h_{125}$, respectively. {\sl Left:} Production cross sections at a 13 \TeV~ LHC. {\sl Right: } Branching ratios of the $h_1\,h_2$ state as a function of the free light scalar mass. The slashed/ hatched/ dotted regions on the benchmark planes are excluded from  comparison with data via \HB/ \HS, the requirement that the potential must be bounded from below, and unitarity constraints. Partially taken from \cite{Robens:2019kga}.}
\end{figure}
\end{center}

In this subsection, I discuss the asymmetric decay modes $h_3\,\rightarrow\,h_1\,h_2$, where successively one of the three scalars is identified with the 125 \GeV~ resonance. I display the corresponding benchmark planes in figure \ref{fig:asymm}.
Depending on the benchmark plane, maximal production cross sections are given by $\sim\,3-4\,\pb,\,\sim\,0.6\,\pb,$ and $0.3\,\pb$ for $h_1\,h_2$ production for BPs 1/2/3, respectively. In BP3, the $h_1h_1h_1$ final state reaches cross sections up to $\sim\,140\,\fb$. Note that as soon as the kinematic threshold for $h_2\,\rightarrow\,h_{125} h_{125}$ is reached, in fact decays from that state become dominant.

Note that the asymmetric BPs in \cite{Robens:2019kga} have been specifically designed such that the $h_1\,h_1\,h_1$ rate is enhanced as soon as the according phase space opens up. This can be in particular observed in the branching ratios for BP3 (bottom right plot in figure 1), where, as soon as $M_2\,\geq\,250\,\GeV$, the $b\,\bar{b}\,b\,\bar{b}\,W^+\,W^-$ final state becomes dominant, surpassing $W^+W^-\,b\,\bar{b}$ despite the phase space and coupling supression. This is however a particular characteristic of this particular benchmark plane.
\subsection{Symmetric decays}
\begin{center}
\begin{figure} 
\begin{center}
\begin{minipage}{\textwidth}
\begin{center}
\includegraphics[width=0.45\textwidth]{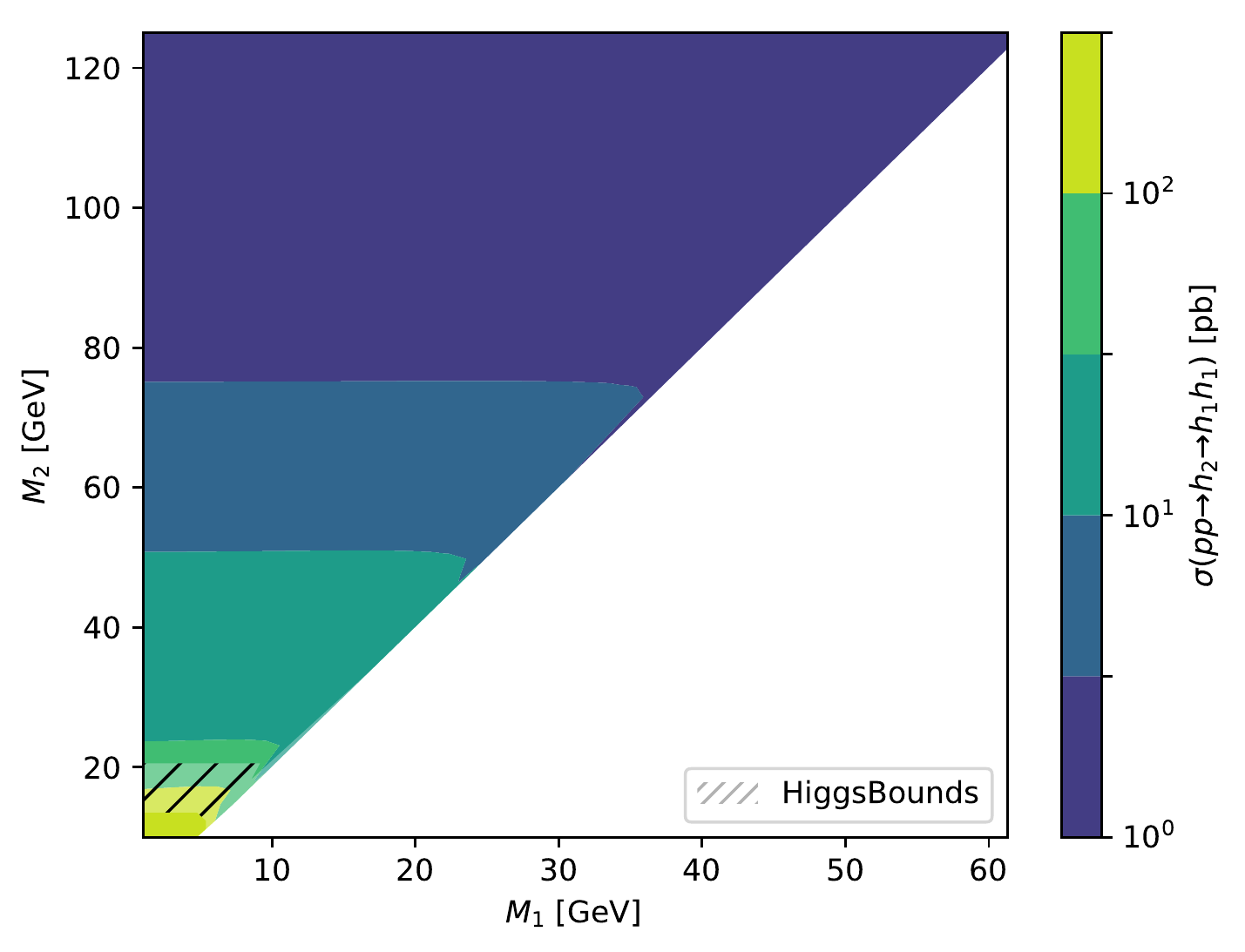}
\includegraphics[width=0.45\textwidth]{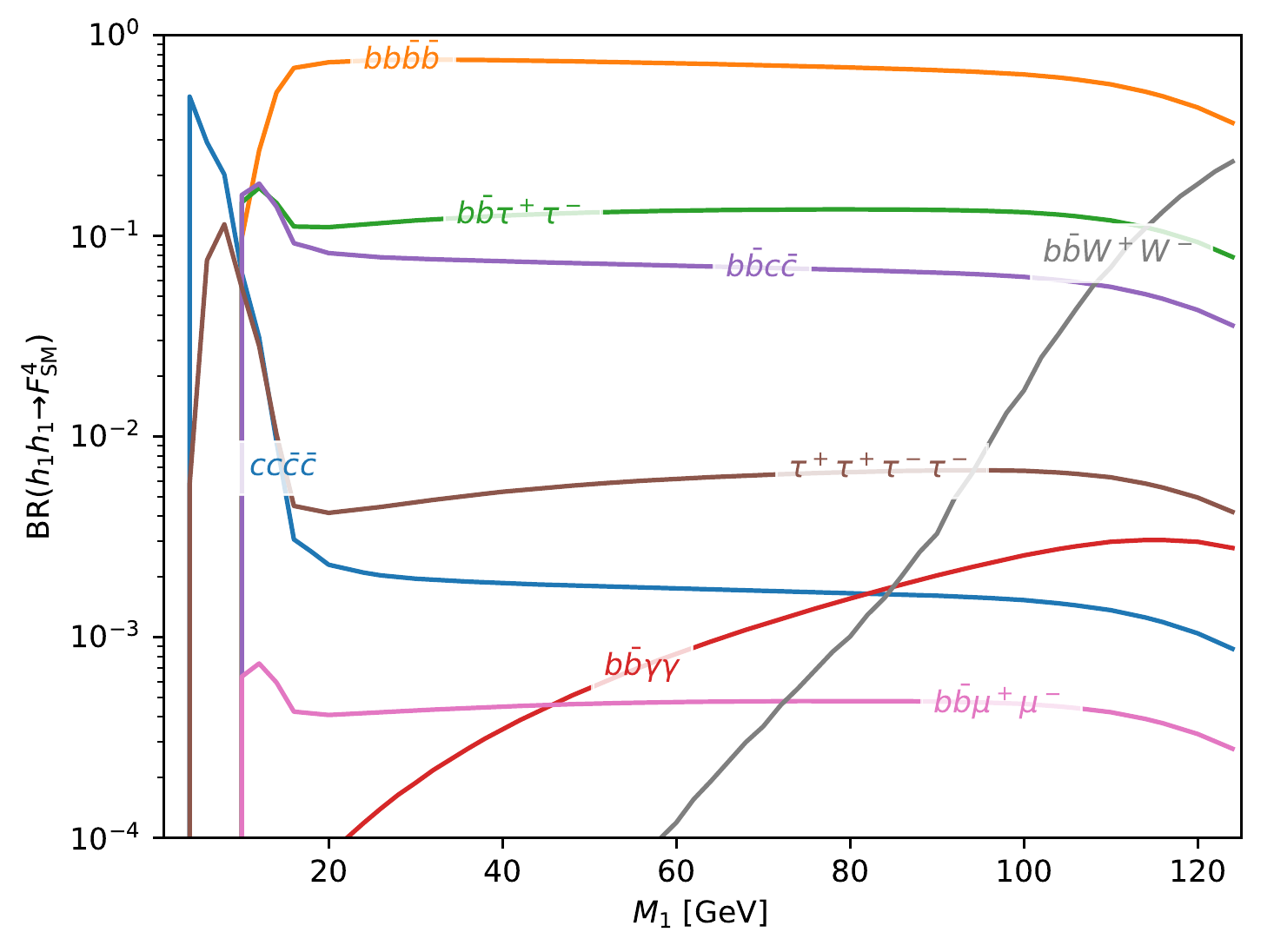}
\\
\vspace{-2mm}
{\small Benchmark Plane 4}
\end{center}
\end{minipage}\\ 
\vspace{2mm}
\begin{center}
\begin{minipage}{0.45\textwidth}
\begin{center}
\includegraphics[width=\textwidth]{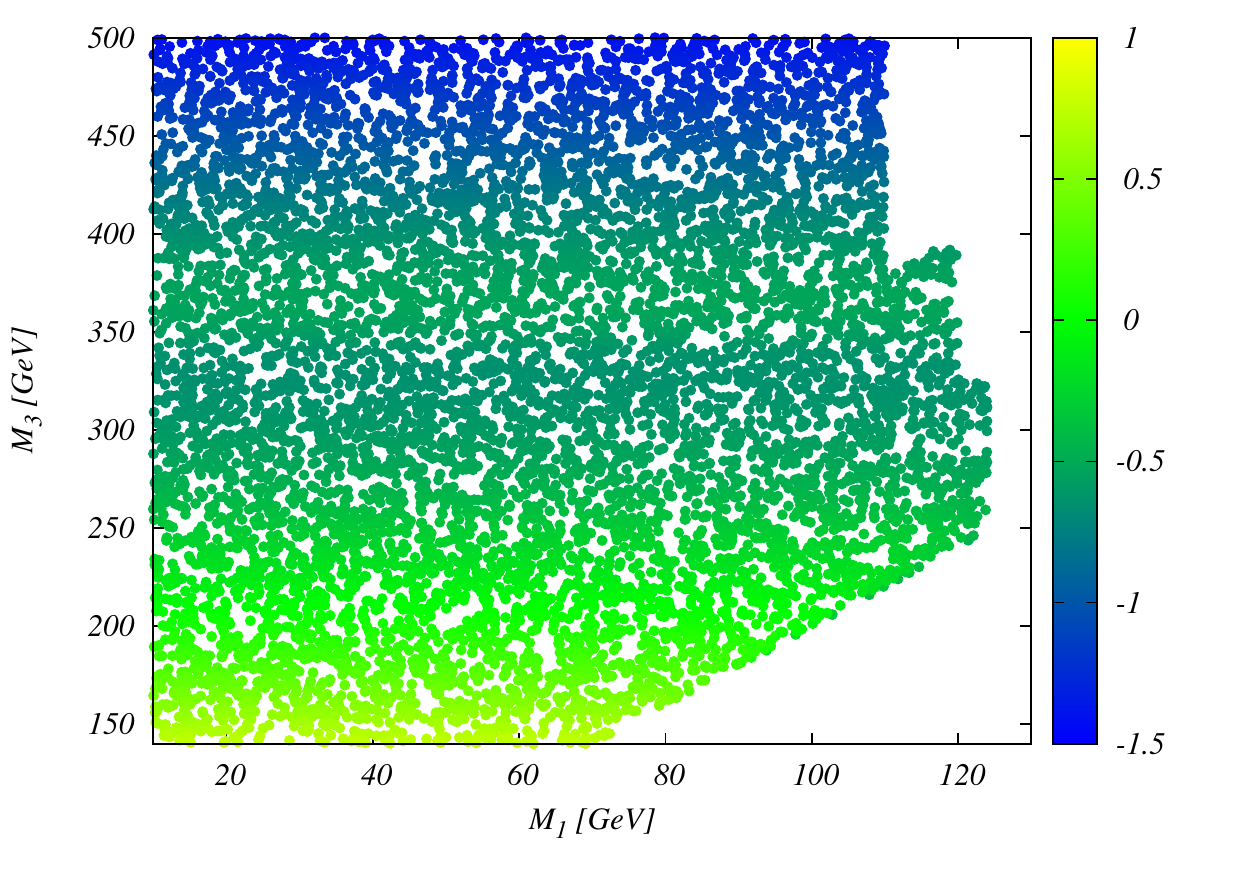}
\end{center}
\end{minipage}
\\
{\small Benchmark Plane 5 [color coding: $\log_{10}\lb \sigma_{h_3\,\rightarrow\,h_1\,h_1}/[\pb]\rb$]}
\end{center}
\vspace{2mm}
\begin{minipage}{\textwidth}
\begin{center}
\includegraphics[width=0.45\textwidth]{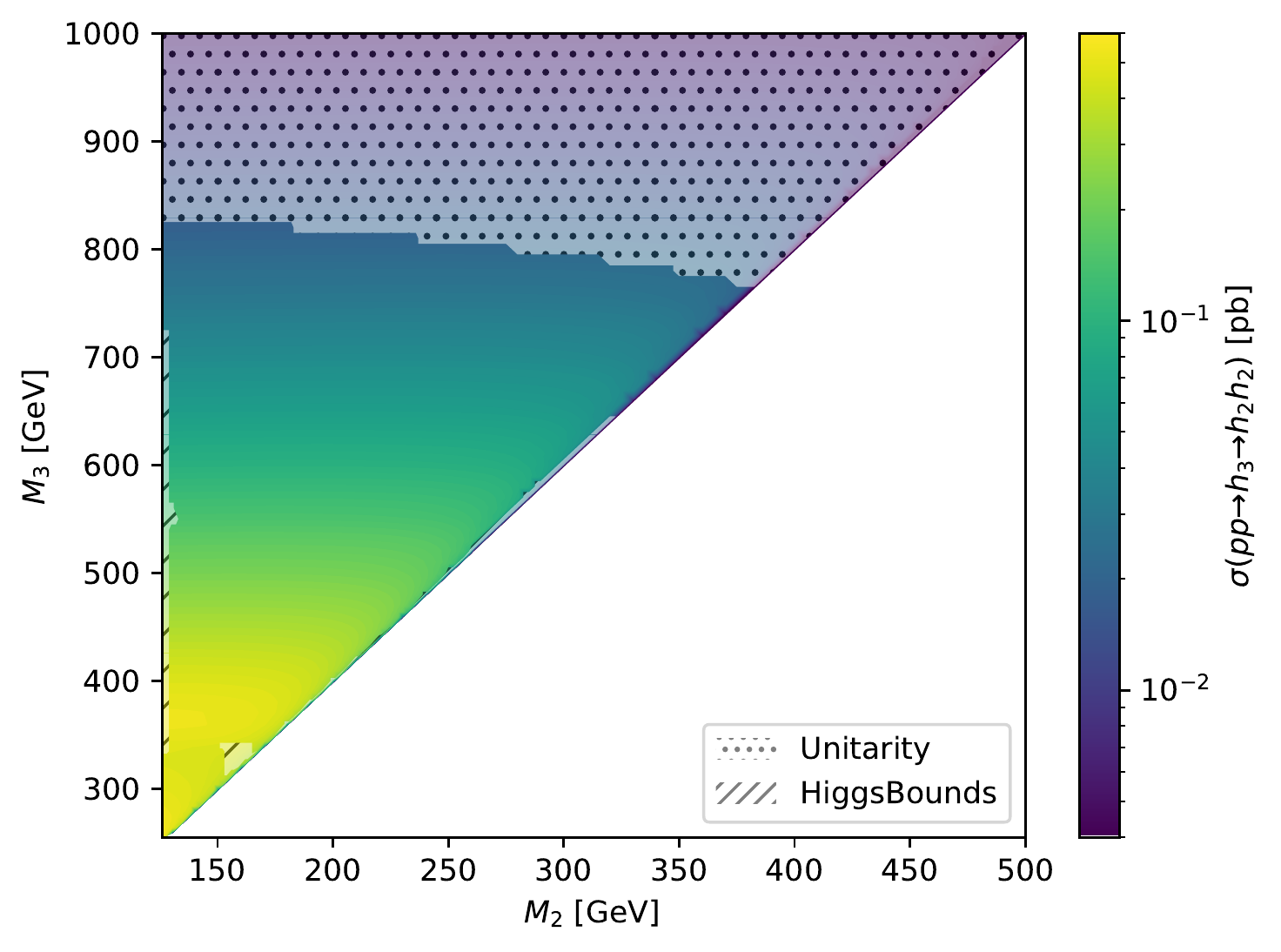}
\includegraphics[width=0.49\textwidth]{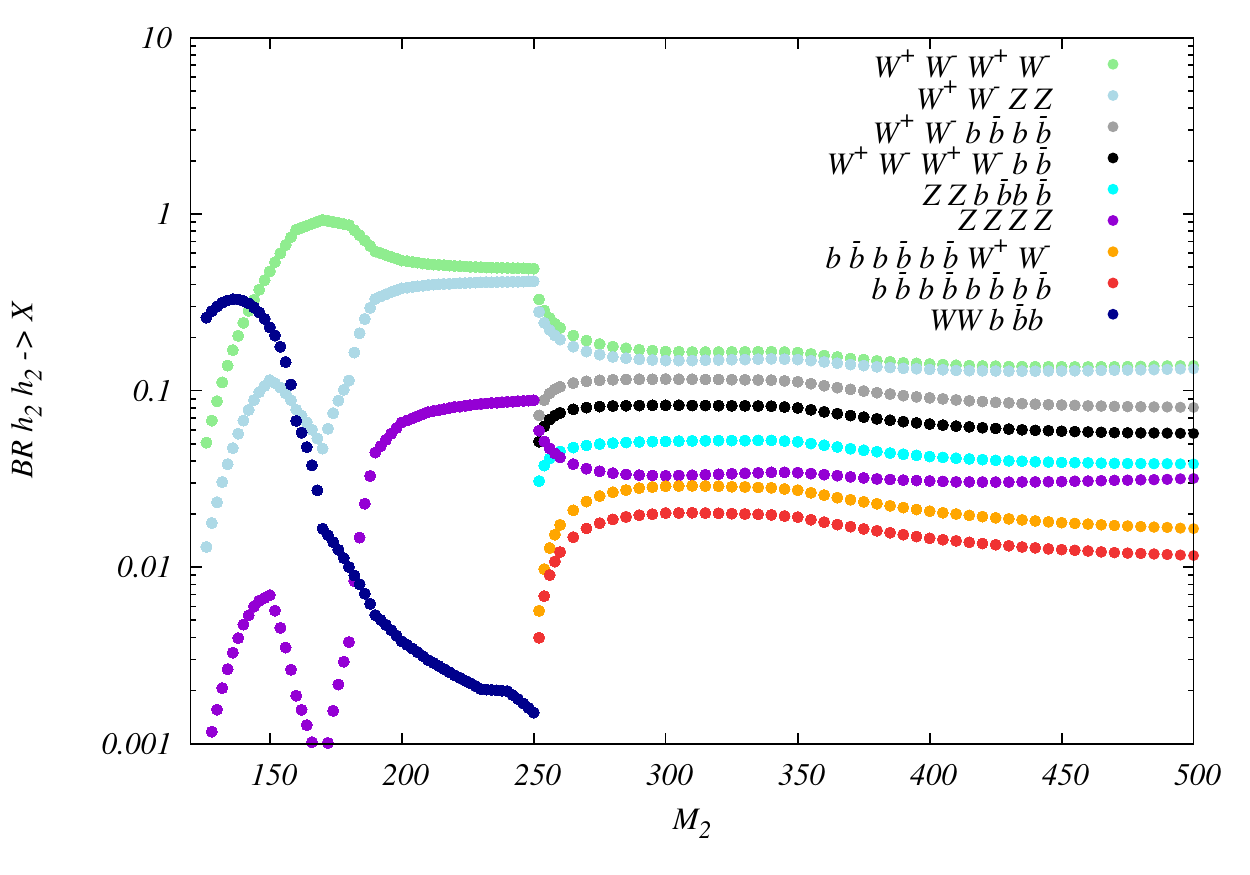}\\
\vspace{-2mm}
{Benchmark Plane 6}
\end{center}
\end{minipage}\\
\end{center}
\vspace{4mm}
\caption{\label{fig:symm}  Benchmark planes for symmetric production and decay, $p\,p\,\rightarrow\,h_{i}\,\rightarrow\,h_j\,h_j$, for various assigments of the 125 \GeV~ resonance. {\sl Top /middle/ bottom rows:} BPs 4/5/6, where $h_{3/2/1}\,\equiv\,h_{125}$. {\sl Left:} Production cross sections at a 13 \TeV~ LHC. {\sl Right: } Branching ratios of the $h_j\,h_j$ state as a function of the lighter free scalar mass. Branching ratios for BP4 and 5 are identical, therefore only one plot is displayed here.  The slashed/ dotted regions on the benchmark planes are excluded from  comparison with data via \HB/ \HS, and unitarity constraints. Partially taken from \cite{Robens:2019kga}.}
\end{figure}
\end{center}

Symmetric decays are given by BPs 4/5/6, with again a differing assignment for $h_{3/2/1}\,\equiv\,h_{125}$, respectively. The corresponding production and decay modes are displayed in figure \ref{fig:symm}.
Depending on the benchmark plane, pair-production cross sections can reach up to 
60/ 2.5/ 0.5 \pb~ for BPs 4/5/6, respectively. For the latter the $h_{125}h_{125} h_{125} h_{125}$ final state can reach rates up to 14 \fb. Also note that the allowed parameter space in BP5 has slightly shrunk, mainly due to the implementation of an additional search \cite{ATLAS:2018rnh} into \HB~ after the performance of the original scan. For BP6, 6 particle final states as e.g. $W^+W^-b\bar{b}b\bar{b}$ can reach branching ratios up to $\sim\,10\%$, depending on $M_2$.

As before, the symmetric benchmark planes in \cite{Robens:2019kga} have been designed to open up for interesting novel final states if the phase space allows for this; for BP6, this means that the $h_2\,\rightarrow\,h_1\,h_1$ rate has been enhanced, reaching up to $40\%$ depending on $M_2$. This again leads to the fact that branching ratios that are dominant prior to the kinematic threshold of $M_2\,\sim\,250\,\GeV$, mainly for electroweak gauge boson final states, are supressed for larger masses. Although they remain dominant, the $W^+\,W^-\,b\,\bar{b}\,b\,\bar{b}$ final state displays similar rates. 
\section{Further investigation of this model}
After the original appearance of the paper proposing the TRSM, several theoretical and experimental works have been performed which at least partially build on the benchmark planes proposed here. We briefly list some of these here.
\subsection{Investigation of the $h_{125}h_{125}h_{125}$ final state}
In BP3, for $M_2\,\rightarrow\, 250\,\GeV$, the decay $h_{2}\,\rightarrow\,h_1\,h_1$ becomes dominant, leading to a $h_{125}h_{125}h_{125}$ final state. For subsequent decays into $b\,\bar{b}$, this BP has been investigated in \cite{Papaefstathiou:2020lyp}. We found that, depending on the parameter point and integrated luminosity, significances between 3 and $\sim\,10$ can be achieved. I display the results in table \ref{table:efficiencies}.

\begin{center}
\begin{figure} 
\begin{center}
  \includegraphics[width=0.6\columnwidth]{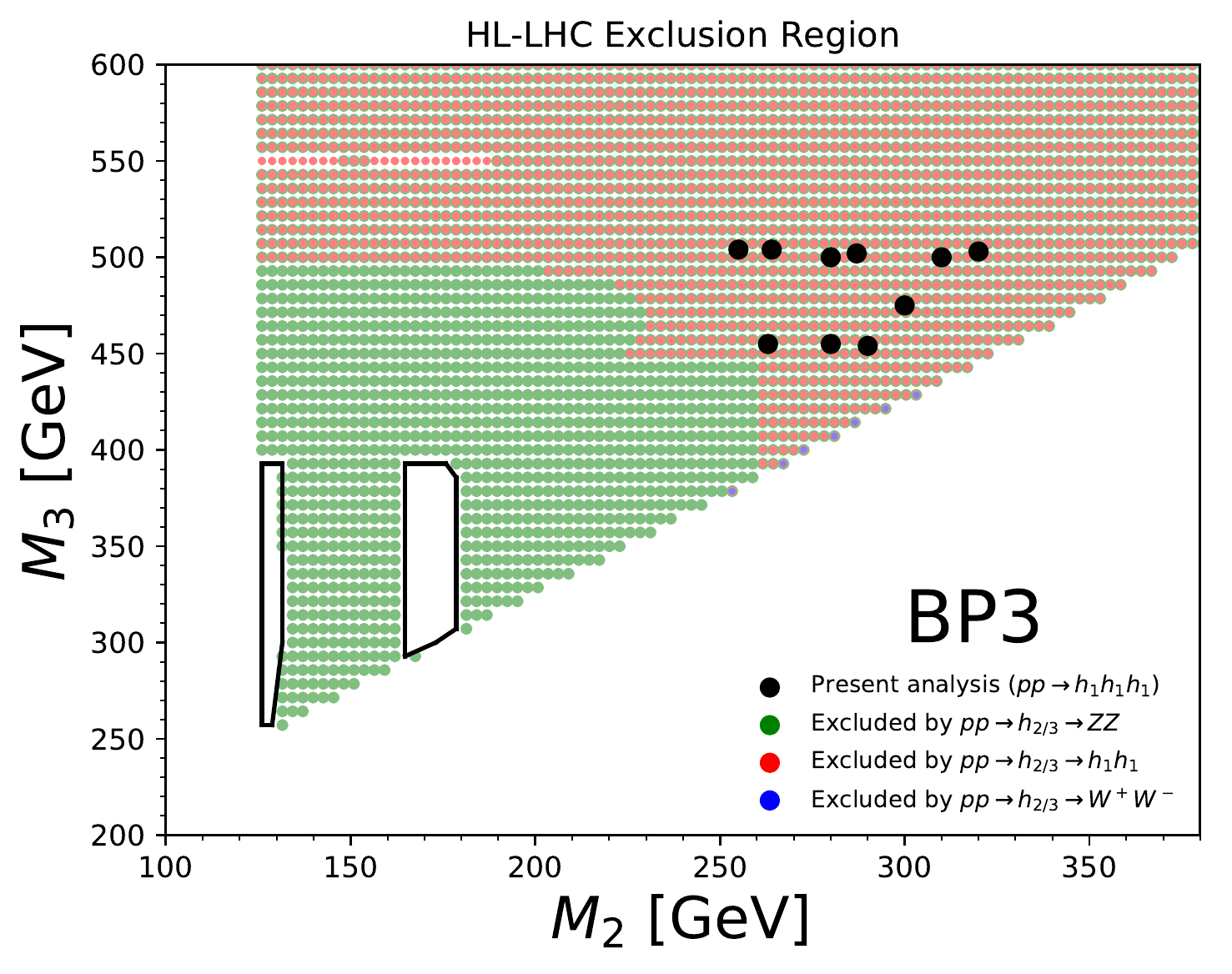}
\caption{\label{fig:const_Andreas} The expected exclusion region for the full integrated luminosity of the HL-LHC, $3000~\rm{fb}^{-1}$, through final states \textit{other} than $p p \rightarrow h_1 h_1 h_1$ as explained in the main text. Points with green circles are expected to be excluded by $ZZ$ final states, with red circles by $h_1 h_1$ and with blue circles by $W^+W^-$. The $W^+W^-$ analysis excludes only very few points on the parameter space and therefore appears infrequently in the figure. The points \textbf{A}--\textbf{I} that we have considered in our analysis of $p p \rightarrow h_1 h_1 h_1$ are shown in black circles overlayed on top of the circles indicating the exclusion. The two cut-out white regions near $M_2 \sim 130$~GeV and $M_2 \sim 170$~GeV will remain viable at the end of the HL-LHC. Taken from \cite{Papaefstathiou:2020lyp}.}
\end{center}
\end{figure}
\end{center}

\begin{table*}[t!]
\begin{tabular*}{\textwidth}{@{\extracolsep{\fill}}cccccccc@{}}
Label&$(M_2, M_3)$ & $\varepsilon_{\rm Sig.}$& $\rm{S}\bigl|_{300\rm{fb}^{-1}}$ & $\varepsilon_{\rm Bkg.}$ & 
$\rm{B}\bigl|_{300\rm{fb}^{-1}}$ & $\text{sig}|_{300\rm{fb}^{-1}}$ & $\text{sig}|_{3000\rm{fb}^{-1}}$\\
& [GeV] & & & & & (syst.) &(syst.)  
\\
\hline
\textbf{A} &$(255, 504)$ & $0.025$ & $14.12$  & $8.50\times 10^{-4}$ & $19.16$ & $2.92~(2.63)$&$9.23~(5.07)$\\
\textbf{B} & $(263, 455)$ & $0.019$ & $17.03$    & $3.60\times 10^{-5}$ & $ {8.12}$ & $4.78 ~(4.50)$&$15.10~(10.14)$\\
\textbf{C} & $(287, 502)$ & $0.030$ & $20.71$ & $9.13\times 10^{-5}$ & $20.60$  & $4.01~(3.56)$ & $12.68~(6.67)$\\
\textbf{D} & $(290, 454)$ & $0.044$ & $37.32$    & $1.96\times 10^{-4}$ & $44.19$& $5.02~(4.03)$&$15.86~(6.25)$\\
\textbf{E} & $(320, 503)$ & $0.051$ & $ {31.74}$    & $2.73\times 10^{-4}$ & $61.55$& $3.76~( {2.87}) $&$11.88~(4.18)$\\
\textbf{F} & $(264,504)$&$0.028$& $18.18$&$9.13\times 10^{-5}$&$20.60$&$3.56~(3.18) $&$11.27~(5.98)$\\
\textbf{G} & $(280, 455)$&$0.044$& $38.70$ &$1.96\times 10^{-4}$& $44.19$ & $5.18~(4.16)$ &$16.39~(6.45)$\\
\textbf{H} & $(300, 475)$ & $0.054$& $41.27$ & $2.95\times 10^{-4}$& $66.46$ & $4.64~(3.47)$&$ 14.68~( {4.94})$\\
\textbf{I} & $(310, 500)$& $0.063$& $41.43$& $3.97\times 10^{-4}$& $89.59$& $4.09~(2.88) $&$ {12.94~(3.87)}$\\
\textbf{J} & $(280,500)$& $0.029 $& $20.67$&$9.14\times 10^{-5}$& $20.60$&$4.00~(3.56) $&$12.65~(6.66)$\\
\end{tabular*}
\caption{ The resulting selection efficiencies, $\varepsilon_{\rm Sig.}$ and $\varepsilon_{\rm Bkg.}$, number of events, $S$ and $B$ for the signal and background, respectively, and statistical significances. A $b$-tagging efficiency of $0.7$ has been assumed. The number of signal and background events are provided at an integrated luminosity of $300~\rm{fb}^{-1}$. Results for $3000~\rm{fb}^{-1}$ are obtained via simple extrapolation. The significance is given at both values of the integrated luminosity excluding (including) systematic errors in the background. Taken from \cite{Papaefstathiou:2020lyp}.}
\label{table:efficiencies}
\end{table*}
Note we also compared how different channels, e.g. direct decays of the heavier scalars into $VV$ or $h_{125}h_{125}$ final states, would perform at a HL-LHC. The results are displayed in figure \ref{fig:const_Andreas}.
We note that all benchmark points that were investigated can additionally be probed by other production and decay mechanisms. Note, however, that these test different regions of the parameter space, as they depend on different parameters in the potential. These searches can therefore be considered to be complementary.

\subsection{Recasting current LHC searches}

\begin{figure} 
\begin{center}
\begin{minipage}{0.45\textwidth}
\includegraphics[width=\textwidth]{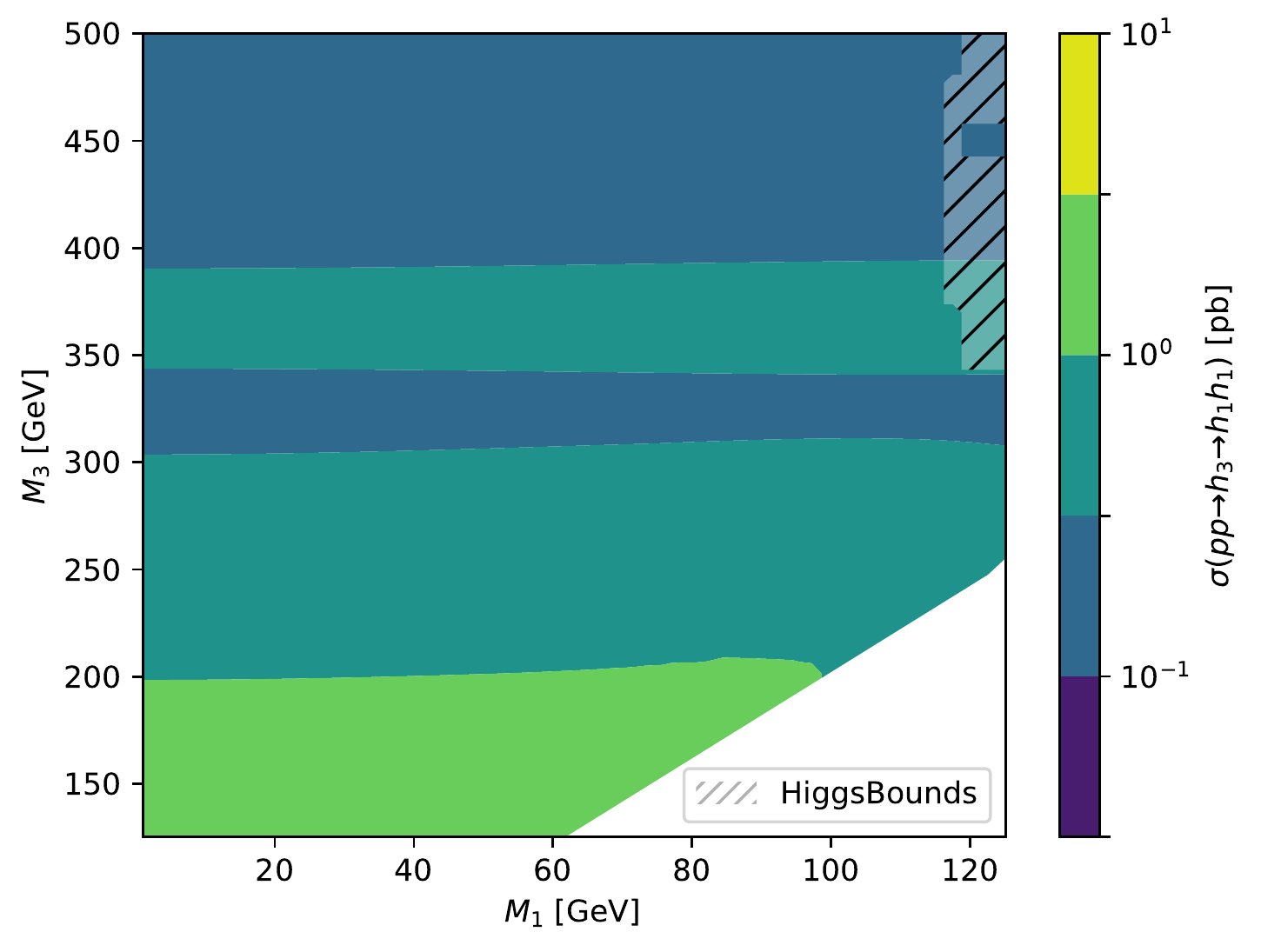}
\end{minipage}
\begin{minipage}{0.45\textwidth}
\includegraphics[width=\textwidth]{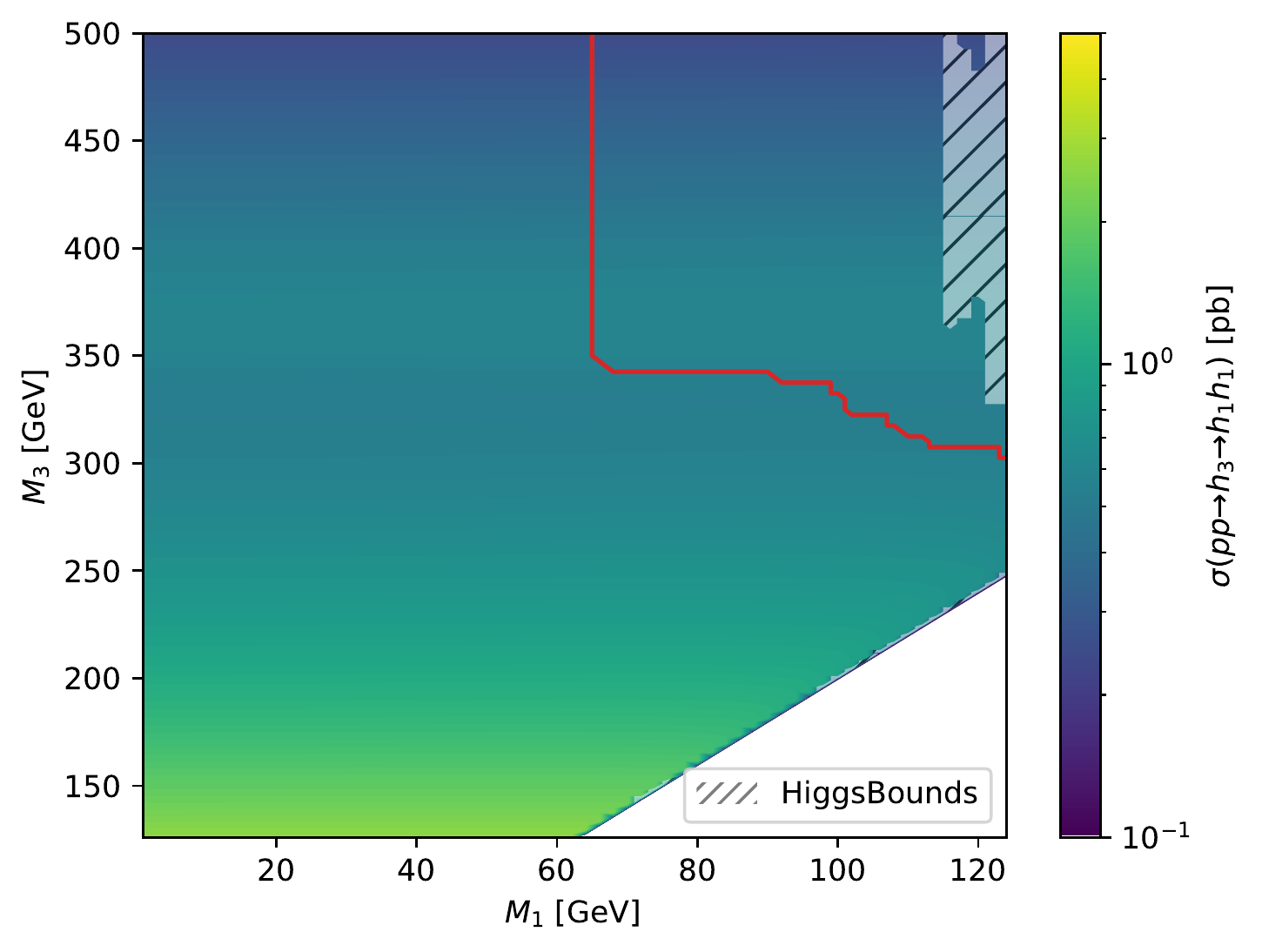}
\end{minipage}
\end{center}
\caption{\label{fig:bp5reint} Reinterpretation of a $36\,\fb^{-1}$ CMS search for di-Higgs production via a heavy resonance using the 4 b final state. The exclusion line uses the results obtained in \cite{Barducci:2019xkq}. Points to the right and above the red contour are excluded.  The slashed regions on the benchmark planes are excluded from  comparison with data via \HB/ \HS. Taken from \cite{Robens:2022mvi}.}
\end{figure}

It is also interesting to investigate whether current searches can be reinterpreted and recasted in such a way that they allow to exclude regions in the models parameter space that were not directly scrutinized in the experimental search, or for which no interpretation was presented in the original publication. In \cite{Barducci:2019xkq}, the authors have reinterpreted a CMS search for $p\,p\,\rightarrow\,H\,\rightarrow\,h_{125}h_{125}\,\rightarrow\,4\,b$ \cite{CMS:2018qmt}, which corresponds to di-Higgs production via a heavy resonance and subsequent decays into $b\,\bar{b}$ final states, and extended the mass ranges for the scalars in the decay chain. I have applied these results to the TRSM, in particular to BP5. I display the corresponding results in figure \ref{fig:bp5reint}\footnote{I thank the authors of \cite{Barducci:2019xkq} for providing us with the corresponding exclusion limits.}. We see that the sensitive region of parameter space is significantly extended, and therefore, an actual experimental analysis also in this parameter region is greatly encouraged.
\subsection{Experimental searches with TRSM interpretations}
\begin{center}
\begin{figure} 
\includegraphics[width=\textwidth]{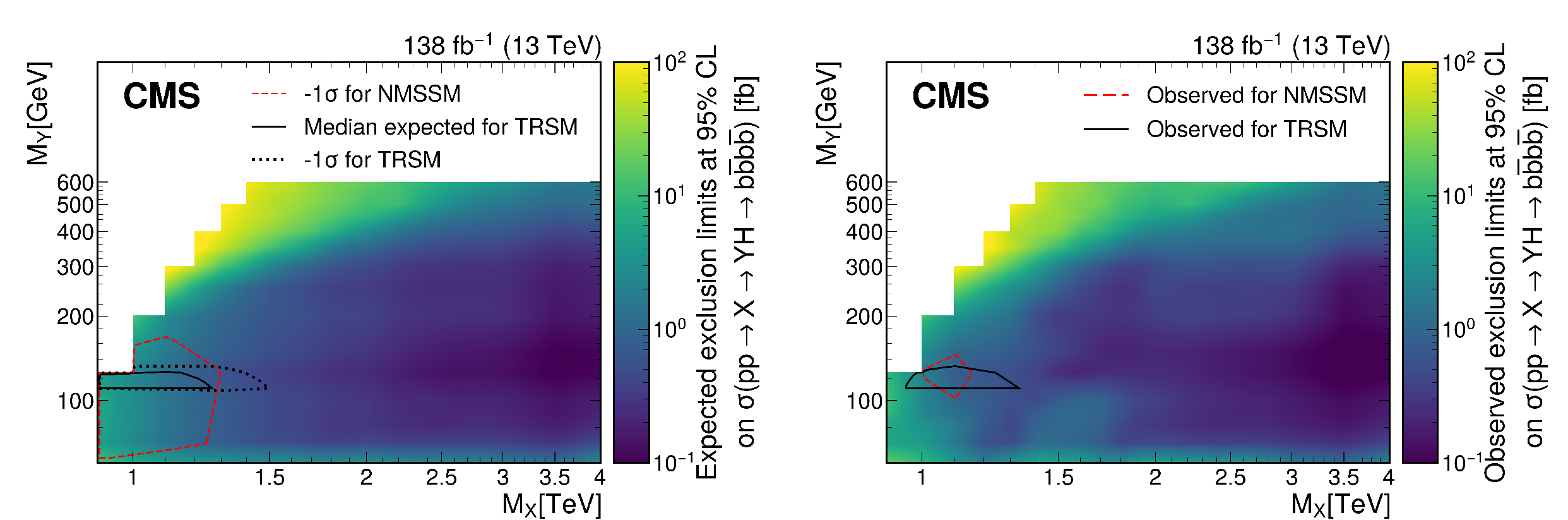}
\caption{\label{fig:cmsres} Expected {\sl (left)} and observed {\sl (right)} $95\%$ confidence limits for the $p\,p\,\rightarrow\,h_3\,\rightarrow\,h_2\,h_1$ search, with subsequent decays into $b\bar{b}b\bar{b}$. For both models, maximal mass regions up to $m_3\,\sim\,\,1.4\TeV,\;m_2\,\sim\,\,140\,\GeV$ can be excluded. Figure taken from \cite{CMS:2022suh}.}
\end{figure}
\end{center}

\begin{center}
\begin{figure} 
\begin{center}
\begin{minipage}{0.45\textwidth}
\begin{center}
\includegraphics[width=\textwidth]{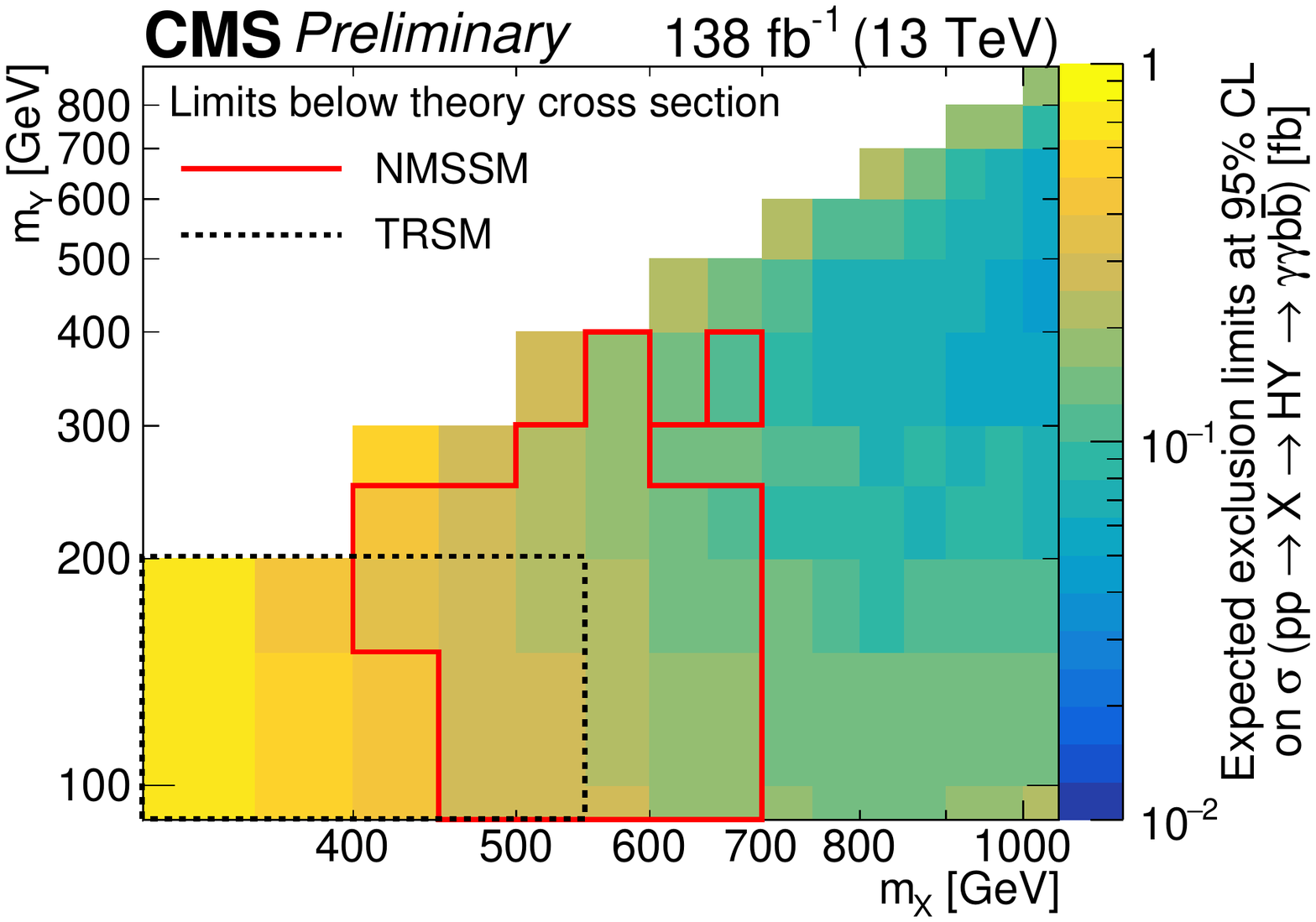}
\end{center}
\end{minipage}
\begin{minipage}{0.45\textwidth}
\begin{center}
\includegraphics[width=\textwidth]{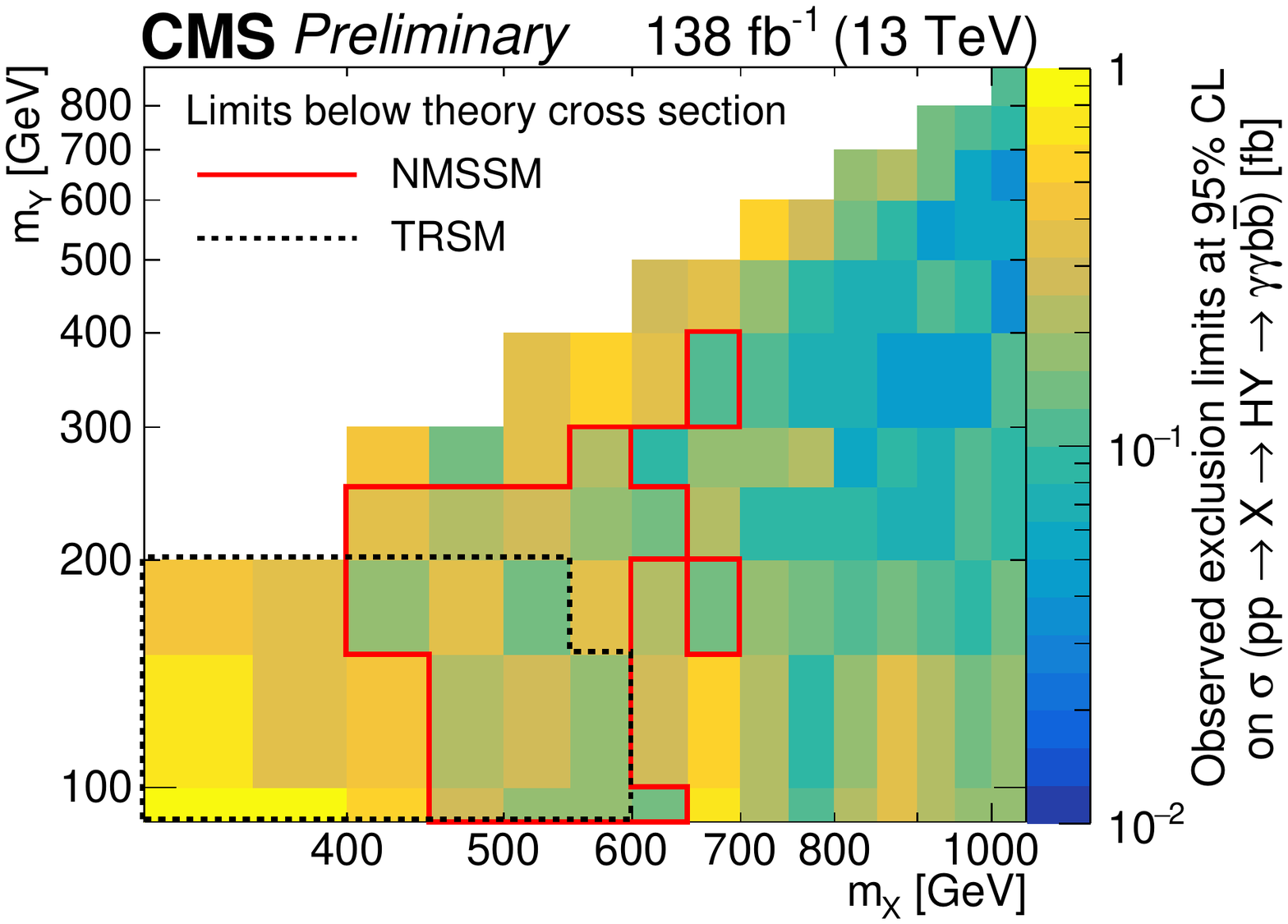}
\end{center}
\end{minipage}
\caption{\label{fig:cmsbbgaga} Expected {\sl (left)} and observed {\sl (right)} $95\%$ confidence limits for the $p\,p\,\rightarrow\,h_3\,\rightarrow\,h_2\,h_1$ search, with subsequent decays into $b\bar{b}\gamma\gamma$. Depending on the model, maximal mass regions up to $m_3\,\sim\,\,800\,\GeV,\;m_2\,\sim\,\,400\,\GeV$ can be excluded. Figure taken from \cite{CMS-PAS-HIG-21-011}.}
\end{center}
\end{figure}
\end{center}

Two experimental searches have by now made use of the predictions obtained within the TRSM to interpret regions in parameter space that are excluded: a CMS search for asymmetric production and subsequent decay into $b\bar{b}b\bar{b}$  final states \cite{CMS:2022suh}, as well as $b\bar{b}\gamma\gamma$ in \cite{CMS-PAS-HIG-21-011}. For this, maximal production cross sections were provided in the parameter space, allowing all additional new physics parameter to float; the respective values have been tabulated in \cite{reptr,trsmbbgaga}. Figures \ref{fig:cmsres} and \ref{fig:cmsbbgaga} show the expected and observed limits in these searches for the TRSM and NMSSM \cite{Ellwanger:2022jtd}.

In addition, several searches also investigate decay chains that can in principle also be realized within the TRSM, as e.g. other searches for the same final states \cite{CMS:2022qww} or $b\,\bar{b}\mu^+\mu^-$ \cite{ATLAS:2021hbr} final states.

\section{Signatures at Higgs factories}
\begin{center}
\begin{figure} 
\begin{center}
\includegraphics[width=0.45\textwidth]{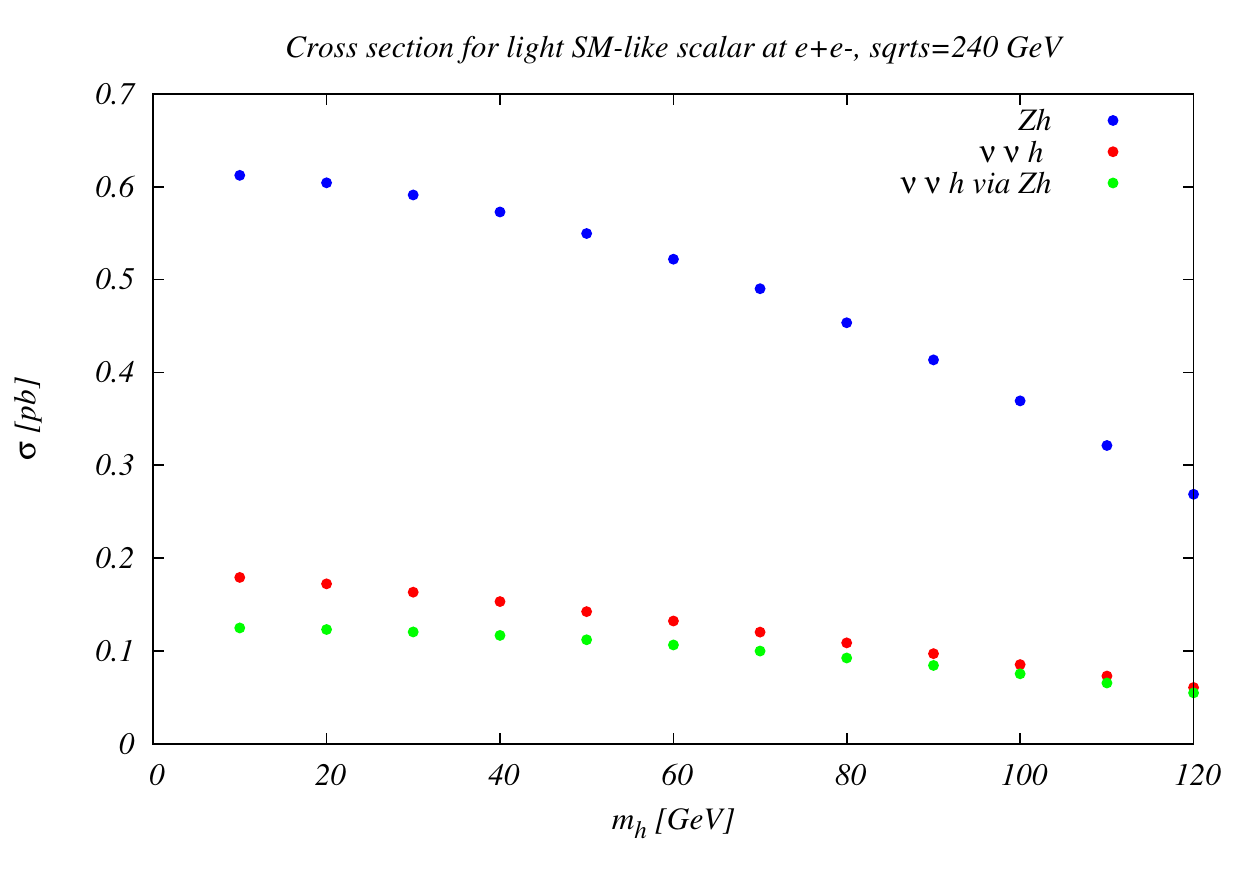}
\includegraphics[width=0.45\textwidth]{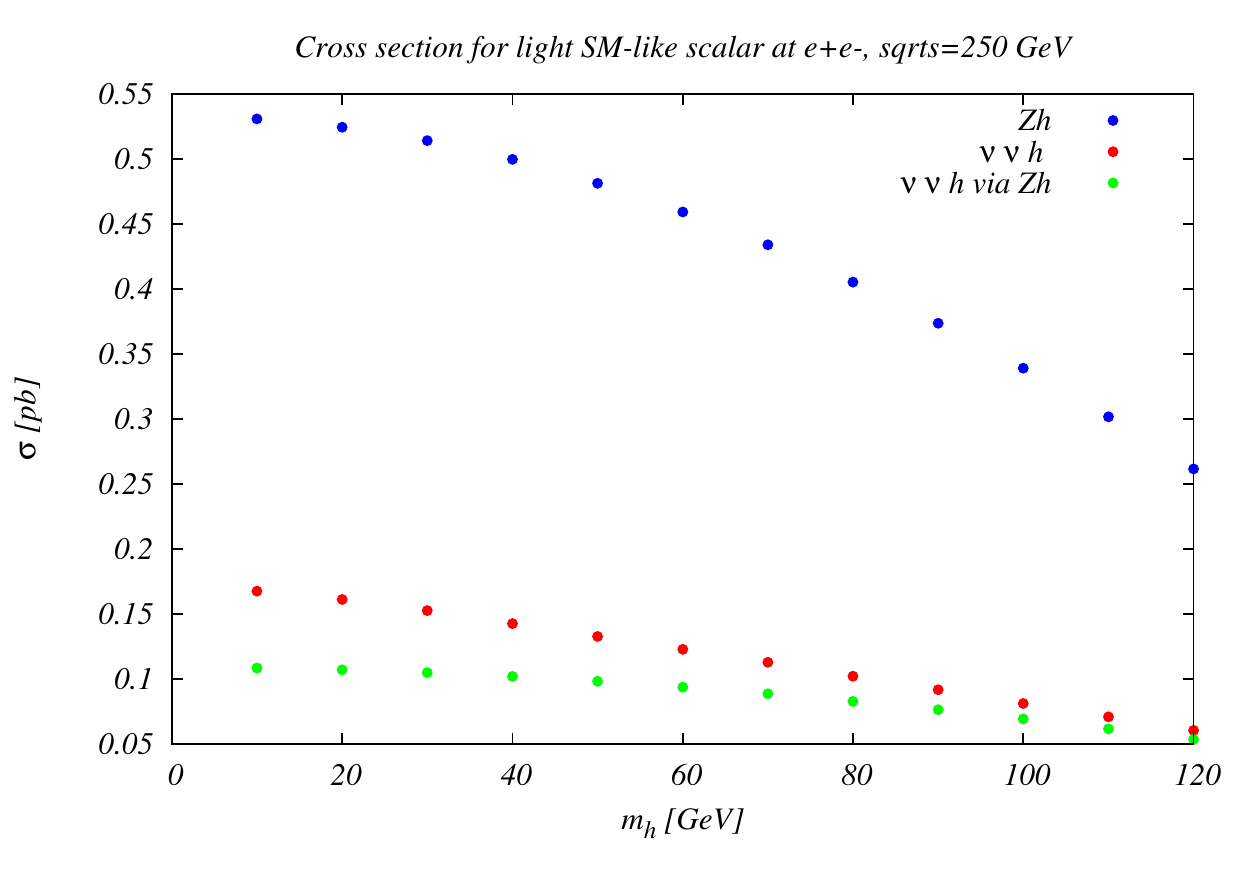}
\caption{\label{fig:prod250} Leading order production cross sections for $Z\,h$ and $h\,\nu_\ell\,\bar{\nu}_\ell$ production at an $e^+\,e^-$ collider with a COM energy of 240 \GeV {\sl (left)} and 250 \GeV~ {\sl (right)} using Madgraph5 for an SM-like scalar h. Shown is also the contribution of $Z\,h$ to $\nu_\ell\,\bar{\nu}_\ell\,h$ using a factorized approach for the Z decay. Taken from \cite{Robens:2022zgk}.}
\end{center}
\end{figure}
\end{center}

The investigation of light scalars has recently gained again more interest, after the recommendation of the European Strategy Report \cite{EuropeanStrategyforParticlePhysicsPreparatoryGroup:2019qin,European:2720129} to concentrate on $e^+e^-$ machines with $\sqrt{s}\,\sim\,240-250\,\GeV$. A short review about the current state of the art for such searches and models which allow for low scalars can e.g. be found in \cite{Robens:2022zgk}. In this model, the only feasible production is $Zh$ radiation of the lighter scalar, with production cross sections given in figure \ref{fig:prod250}. Cross sections have been derived using Madgraph5 \cite{Alwall:2011uj}.

We can now investigate what would be production cross sections for scalar particles with masses $\lesssim\,160\,\GeV$ at Higgs factories.
\subsection{Production of 125 \GeV~ resonance and subsequent decays}
We first turn to the easy case of the production of the 125 \GeV~ resonance in various benchmark scenarios. Of interest are cases where decays $h_{125}\,\rightarrow\,h_i\,h_j$ are kinematically allowed. Note that our benchmark points were not set up in particular for the scenario where $i\,=\,j$, so for this rates might be relatively small by construction.

From table \ref{tab:BPparams} we see that for all scenarios the rescaling for the 125 \GeV~ resonance is $\gtrsim\,0.966$, leading to production cross sections of about $\sim\,0.2\,\pb$, close to the SM value. In general, due to constraints from the invisible branching ratio \cite{ATLAS-CONF-2020-052} as well as signal strength fits, the production cross section for $h_i\,h_j$ final states has to be lower by at least an order of magnitude, leading to cross sections $\mathcal{O}\lb 10\,\fb\rb$. In fact, in the benchmark planes presented here the largest rate for $Z\,h_{125}$ production and subsequent scalar decays can be found in BP1, where the rates are given by multiplying the BRs from figure \ref{fig:asymm} with the production of $Z\,h_{125}$, giving maximal cross sections of around 18 \fb.

\subsection{Additional scalar production}
\begin{center}
\begin{figure} 
\begin{center}
\includegraphics[width=0.45\textwidth]{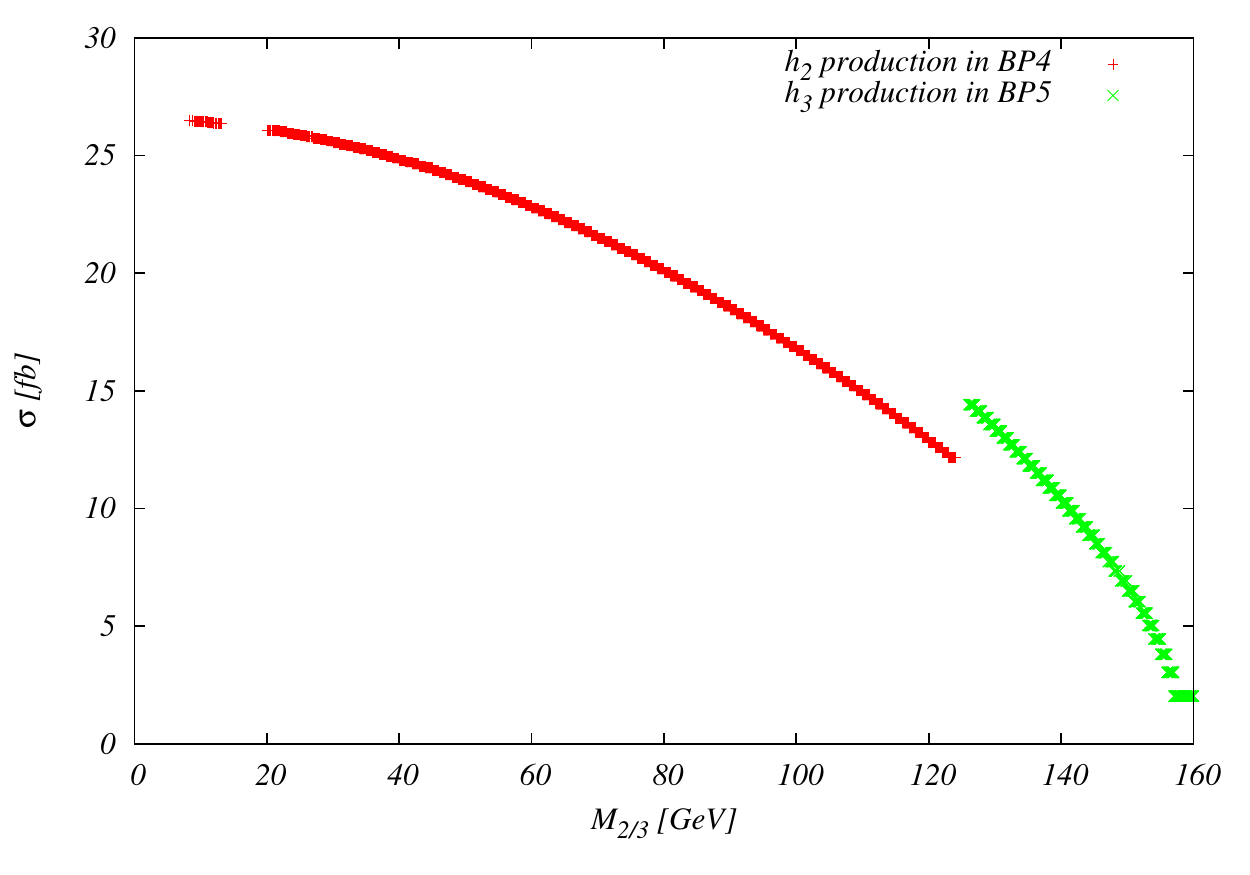}
\end{center}
\caption{\label{fig:prod45} Production cross sections for $Z h_{2/3}$ in BPs 4 and 5, respectively, at a 250 \GeV~ Higgs factory.}
\end{figure}
\end{center}

We now turn to the Higgs-Strahlung production of new physics scalars. This process is in principle possible in all BPs discussed here. However, if we require production rates of $Z\,h_i$ to be larger than $\sim\,10\,\fb$, only BPs 4 and 5 render sufficiently large rates for the production of $h_2$ and $h_3$, respectively. Production rates are independent of the other scalars, and we therefore depict them for both BPs in figure \ref{fig:prod45}. Note that BP4 and BP5 have slightly different parameter settings, in particular the absolute value of $\kappa_3\,=\,-0.250$ in BP5 is slightly larger than the absolute value of $\kappa_2\,=\,0.223$ in BP4, leading to a discontinuity for the production cross section predictions in that figure.
BP4 is constructed in such a way that as soon as the corresponding parameter space opens up, the $h_1\,h_1$ decay becomes dominant; final states are therefore mainly $Z\,b\bar{b}b\bar{b}$ if $M_2\,\gtrsim\,2\,M_1$. Below that threshold, dominant decays are into a $b\,\bar{b}$ pair, which means that standard searches as e.g. presented in \cite{Drechsel:2018mgd,Wang:2020lkq} should be able to cover the parameter space.

Similarly, in BP5 the $h_3\,\rightarrow\,h_1\,h_1$ decay is also favoured as soon as it is kinematically allowed. Therefore, in this parameter space again $Z b\bar{b}b\bar{b}$ final states become dominant. Otherwise $Z\,b\bar{b}$ and $ZW^+W^-$ final states prevail, with a cross over for the respective final states at around $M_3\,\sim\,135\,\GeV$. Branching ratios for these final states are in the $40-50\%$ regime.

\section{More general scan}
So far, I have constrained myself to the discussion of the benchmark planes which were presented in \cite{Robens:2019kga}. However, of course it is also of interest to consider generic scans of the model, and/ or other parameter regions. An example for this has already been given above, where a more generic parameter region was investigated in \cite{CMS:2022suh,reptr}.

Here, I plan to concentrate on scenarios that are accessible at future Higgs factories. One reason for this is that while the BPs in \cite{Robens:2019kga} were especially designed to focus on by that time non-explored signatures at the LHC, the production and decay processes at lepton colliders are slightly more constrained, as stated above. Second, the inclusion of additional low-mass scalars might help to reduce the discrepancy between the SM prediction and experimental PDG value of the W-boson mass, see e.g. an early discussion in \cite{Lopez-Val:2014jva} in the context of a real singlet extension.

\begin{center}
\begin{figure}
\includegraphics[width=0.45\textwidth]{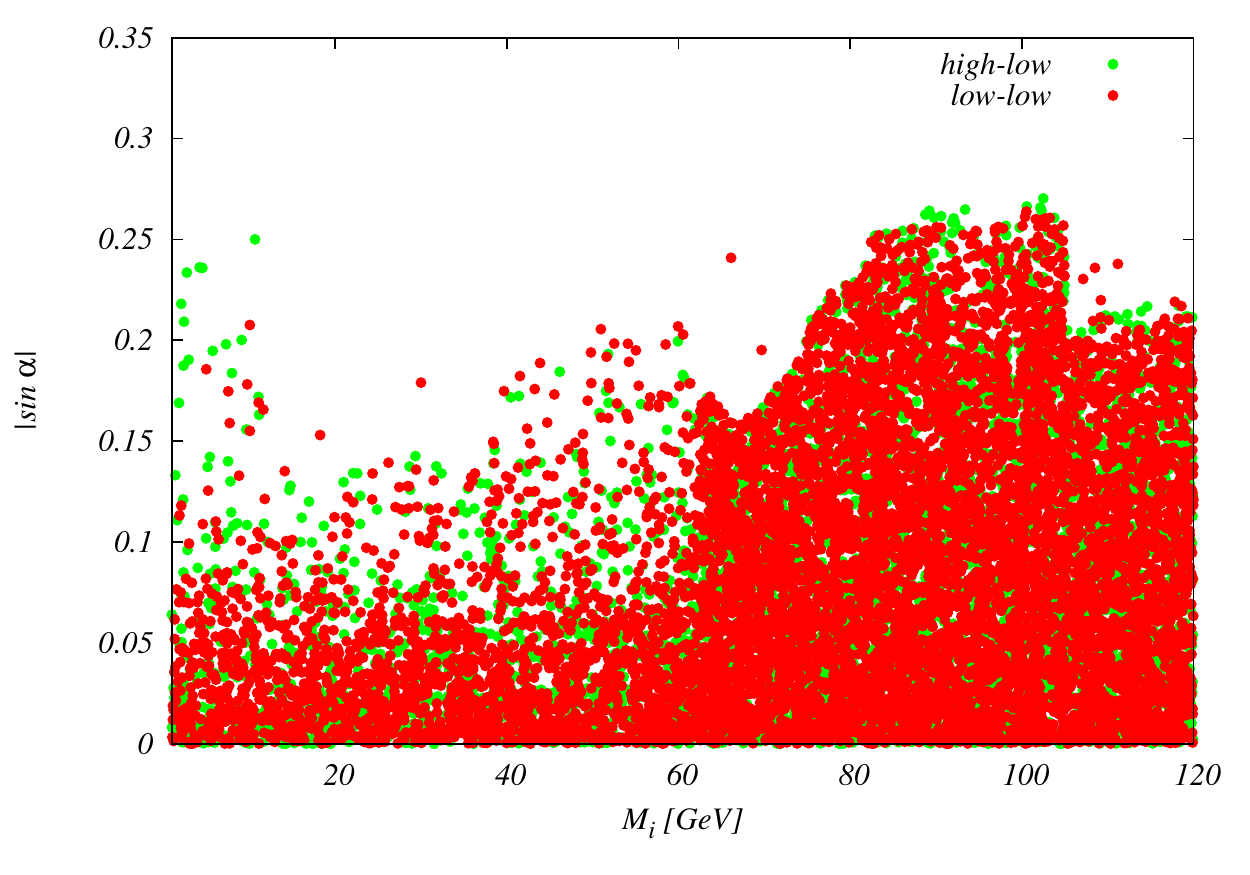}
\includegraphics[width=0.45\textwidth]{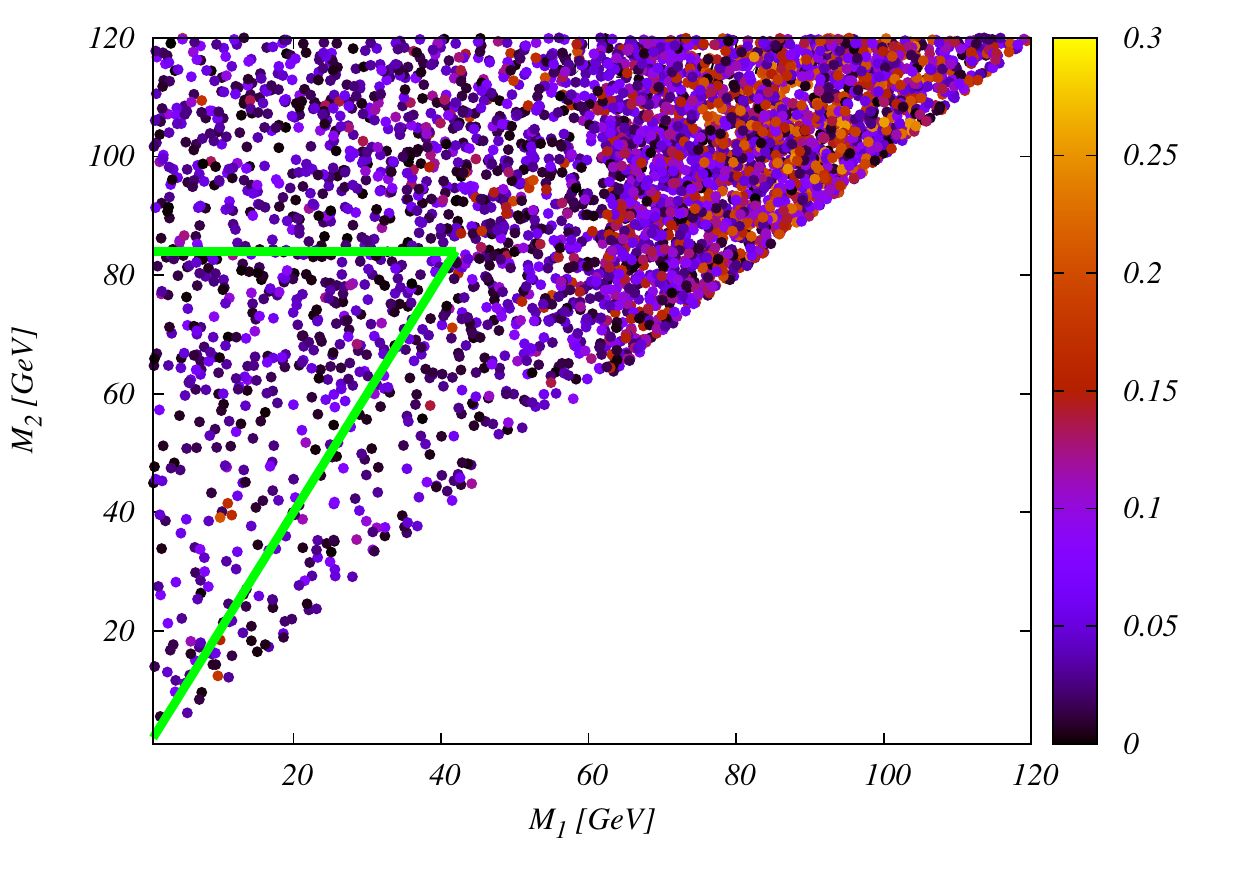}
\caption{\label{fig:trsm} Available parameter space in the TRSM, with one (high-low) or two (low-low) masses lighter than 125 \GeV. {\sl Left}: light scalar mass and mixing angle, with $\sin\al\,=\,0$ corresponding to complete decoupling. {\sl Right:} available parameter space in the $\lb M_{1},\,M_{2}\rb$ plane, with color coding denoting the rescaling parameter $\sin\al$ for the lighter scalar $h_1$. Within the green triangle, $h_{125}\,\rightarrow\,h_2 h_1\,\rightarrow\,h_1\,h_1\,h_1$ decays are kinematically allowed. Taken from \cite{Robens:2022zgk}.}
\end{figure}
\end{center}

I start with presenting the general result of a scan in the 2 mass or 1 mass 1 mixing angle plane already given in \cite{Robens:2022erq,Robens:2022zgk}, given in figure \ref{fig:trsm}. In this figure, two data-sets are considered which fulfill all current constraints as implemented using the current versions of \texttt{ScannerS} and \HB, \HS. The are labelled "low-low" if both $M_{1,2}\,\leq\,125\,\GeV$ and "high-low" if $M_1\,\leq\,125\,\GeV,\,M_3\,\geq\,125\,\GeV$.

In that plot, $|\sin\al|$ is symbolic for the respective mixing angle, earlier denoted by $\kappa_i$, where $\sin\al\,=\,0$ would correspond to the complete decoupling. We see that in general, for low mass scalars, mixing angles up to $\sim\,0.3$ are still allowed. This also in principle can lead to slightly higher production rates than discussed in the previous section.

\subsection{W-boson mass in the TRSM}

In general, for extensions of the scalar sector by one or several gauge-singlets, the contributions to the W-boson mass can be factorized into a SM and a new physics part, as discussed in \cite{Lopez-Val:2014jva} for a real singlet extension. The extension of this for an additional singlet is straightforward, leading to the following expression of $\Delta \lb \delta \rho_{\text{TRSM}}\rb$:
\begin{\eqn*}
\Delta \lb \delta \rho_{\text{TRSM}}\rb\,=\,\sum_i \Delta \lb \delta \rho_{\text{sing}}\rb \lb M_{i}; \kappa_i\rb
\end{\eqn*}
where  $\Delta \lb \delta \rho_{\text{sing}}\rb \lb M_{i}; \sin\al_i\rb$ is given by eqn. (26) of \cite{Lopez-Val:2014jva} with the replacement $m_{H^0}\,\rightarrow\,M_i,\,\sin\al\,\rightarrow\,\kappa_i$, and $M_i\,\neq\,125\,\GeV$ in the above sum. The relation $\sum_i \kappa_i^2\,=\,1$ ensures in fact that the above relation holds in general for a model with an arbitrary number of singlet extensions.

In the comparison with the current measurement of the W-boson mass \cite{ParticleDataGroup:2022pth},

\begin{\eqn*}
M_W^\text{exp}\,=\,\lb 80.377\,\pm\,0.012\rb\,\GeV,
\end{\eqn*}

we have also updated the input values for the SM prediction, as already presented in \cite{Papaefstathiou:2022oyi}, i.e. we use

\begin{eqnarray*}
\al_s\lb M_Z\rb\,=\,0.1179;&M_h\,=\,125.25\,\GeV;&M_t\,=\,172.76\,\GeV;\\
M_Z\,=\,91.1876;&\Delta \al_\text{had}\,=\,276\,\times\,10^{-4};&\Delta \al_\text{lep}\,=\,314.979\,\times\,10^{-4} \;,
\end{eqnarray*}
which gives $M_W^\text{SM}\,=\,80.356\,\GeV$ as the SM prediction, following the calculation outlined in \cite{Awramik:2003rn}.

\begin{center}
\begin{figure}
\begin{center}
\includegraphics[width=0.5\textwidth]{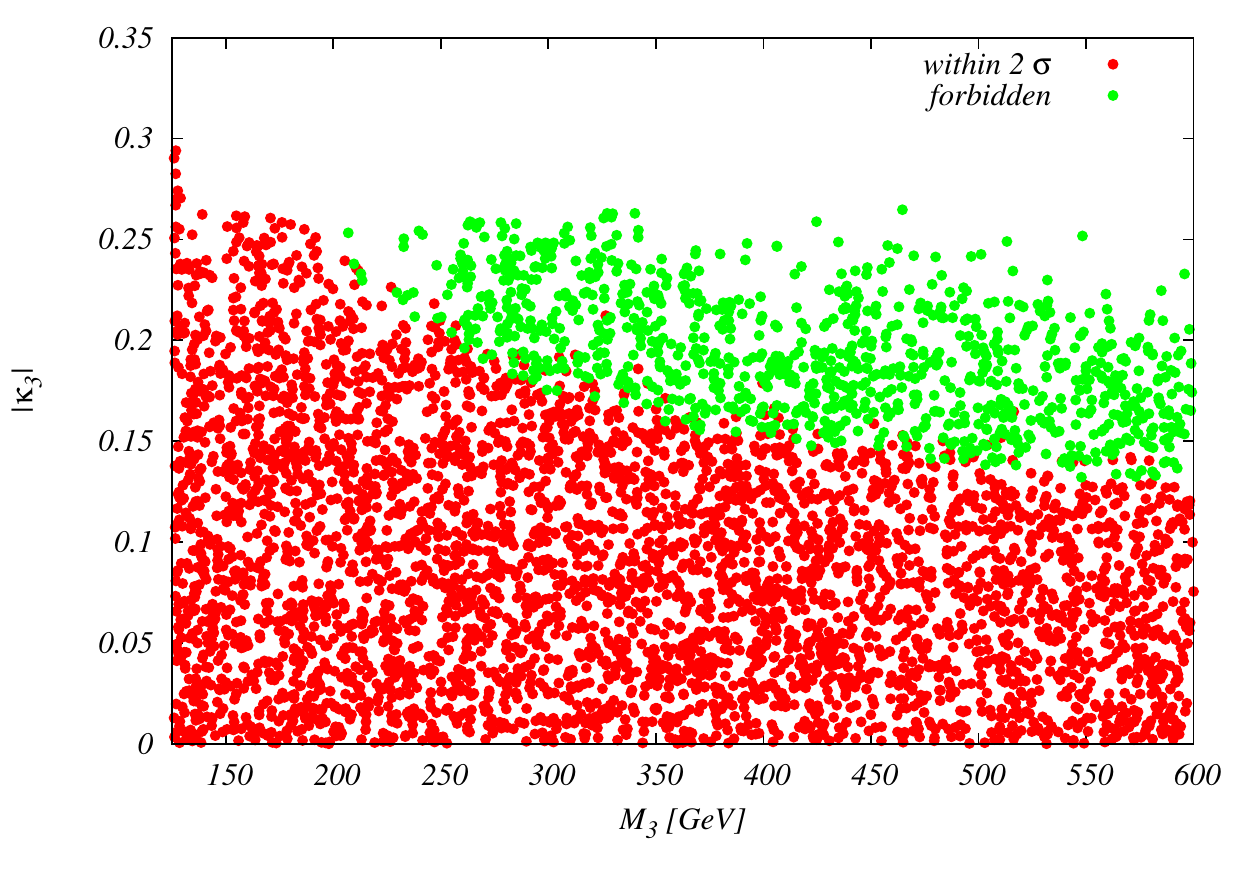}
\caption{\label{fig:mwtrsmh} Allowed (red) and excluded (green) regions in the $\lb M_3;|\kappa_3| \rb$ plane for the "high-low" dataset, where $M_3\,\gtrsim\,M_{125}$. Regions roughly above $M_3\,\gtrsim\,200\,\GeV,\,|\kappa_3|\,\gtrsim\,0.15$ can be excluded requiring a maximal 2 $\sigma$ discrepancy between prediction and experimentally allowed value.}
\end{center}
\end{figure}
\end{center}

We then evaluate the new physics contributions to the W-boson mass by extending the code presented in \cite{Lopez-Val:2014jva} by contributions from a second scalar, where the mass is determined recursively as discussed in that work, and compare it to the current experimental value given above, requiring an at most 2 $\sigma$ discrepancy.
As expected, for the "low-low" dataset introduced above where $M_3\,\equiv\,M_{125}$, corrections drive the W-boson mass prediction closer to the SM, so none of the points is excluded by requiring a maximal 2 $\sigma$ discrepancy. On the other hand, for the "high-low" dataset, where $M_2\,\equiv\,M_{125}$, points with masses and mixing angles $m_3\,\gtrsim\,200\,\GeV,\,|\kappa_3|\,\gtrsim\,0.15$ can be ruled out, cf. figure \ref{fig:mwtrsmh}, where the maximally allowed mixing angle is mass-dependent.
Finally, one can ask whether the current $\sim\,1.8\,\sigma$ discrepancy between experimental value and SM prediction can be significantly reduced within the TRSM taking new physics contributions into account. In general, this would require relatively light masses, together with large mixing angles for such masses. In the datasets investigated here, the maximal value for the W-boson mass was around $M_W\,\sim\,80.361\,\GeV$. This is in fact a point in the high-low dataset, where however the heavier scalar is nearly decoupled. The exact input parameters for this point are given by
\begin{eqnarray*}
&&M_1\,=\,4.2\,\GeV,\,M_3\,=\,494\,\GeV,\;\kappa_1\,=\,0.24,\,\kappa_3\,=\,0.016.
\end{eqnarray*}
In general, scenarios with lightest scalars with masses $M_1\,\lesssim\,12\,\GeV,\,|\kappa_1|\,\gtrsim\,0.15$ give the largest positive corrections to the W-boson mass. Several of such points exist in both the high-low and low-low datasets, cf figure \ref{fig:trsm} (left). The above discussion also shows that taking into account all current constraints, the TRSM cannot explain even larger deviations for the W-boson mass, as e.g. the values reported in \cite{CDF:2022hxs} that range from $80.433\,\pm\,9\,\GeV$ to $80.424\,\pm\,9\,\GeV$ for single measurement and combination, respectively.
\subsection{Production cross sections at a Higgs factory}
Finally, we investigate the maximal allowed production cross section at Higgs factories, where, as before, we chose $\sqrt{s}\,=\,250\,\GeV$ as a benchmark center-of-mass energy. As discussed above, $Z\,h$ production is dominant in the low mass range, and also gives the largest contribution to the $\nu\bar{\nu}h$ final state, so we concentrate on Higgs-strahlung.

\begin{center}
\begin{figure}
\begin{center}
\includegraphics[width=0.49\textwidth]{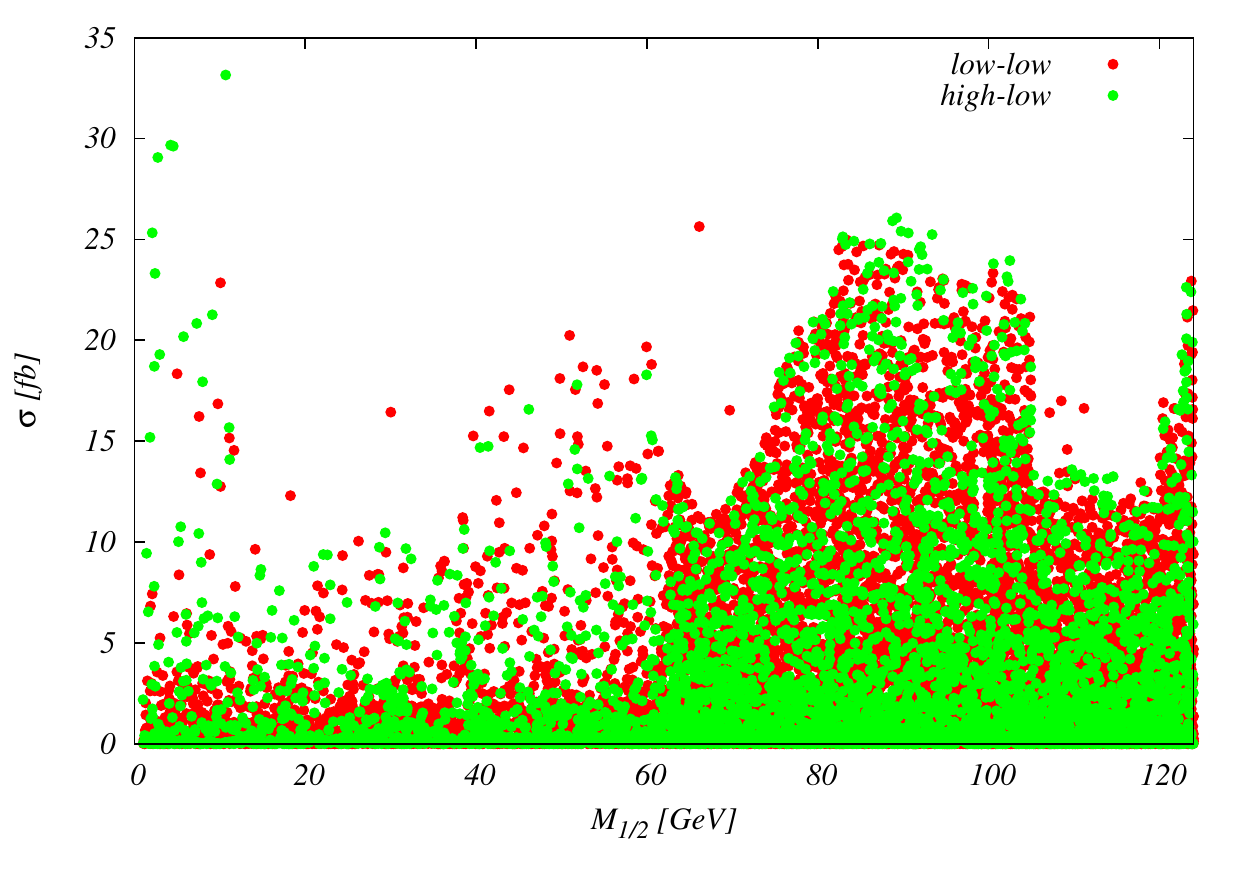}
\includegraphics[width=0.49\textwidth]{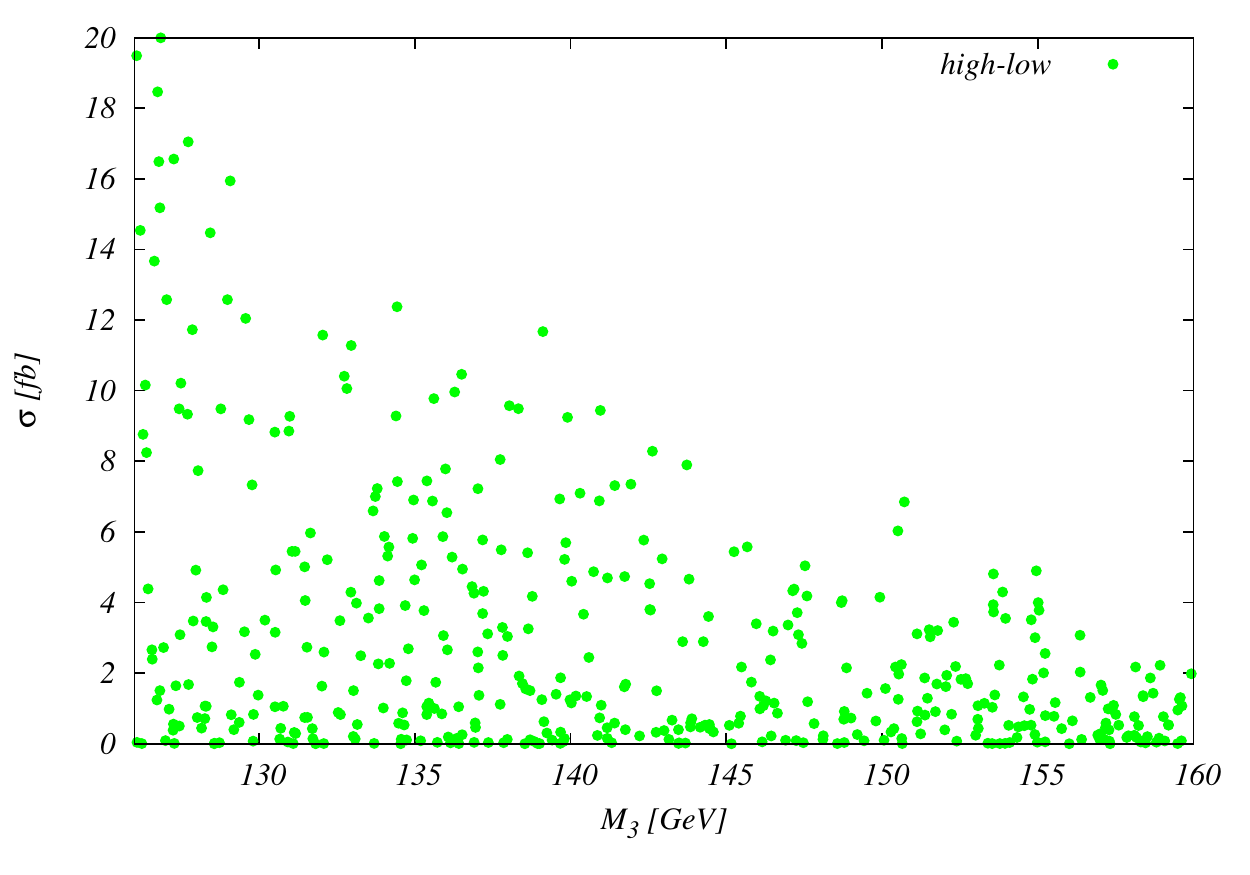}
\caption{\label{fig:xsecs} Maximal production cross section for Higgs-Strahlung for scalars of masses $\neq\,125\,\GeV$ in the TRSM for points passing all discussed constraints. Production cross sections depend on the parameter point and can reach up to 30 \fb.}
\end{center}
\end{figure}
\end{center}

We show maximally allowed production cross sections at an $e^+e^-$ collider with a COM energy of 250 \GeV~ in figure \ref{fig:xsecs}. Note we do not display the region where $M_i\,\sim\,125\,\GeV$; here, when the other scalars are close to being decoupled (in the sense that $|\kappa_i|\,\sim\,0$), we recover production cross sections around 250 \fb, as predicted for the SM using the LO approach discussed here.

\begin{center}
\begin{figure}
\begin{center}
\includegraphics[width=0.5\textwidth]{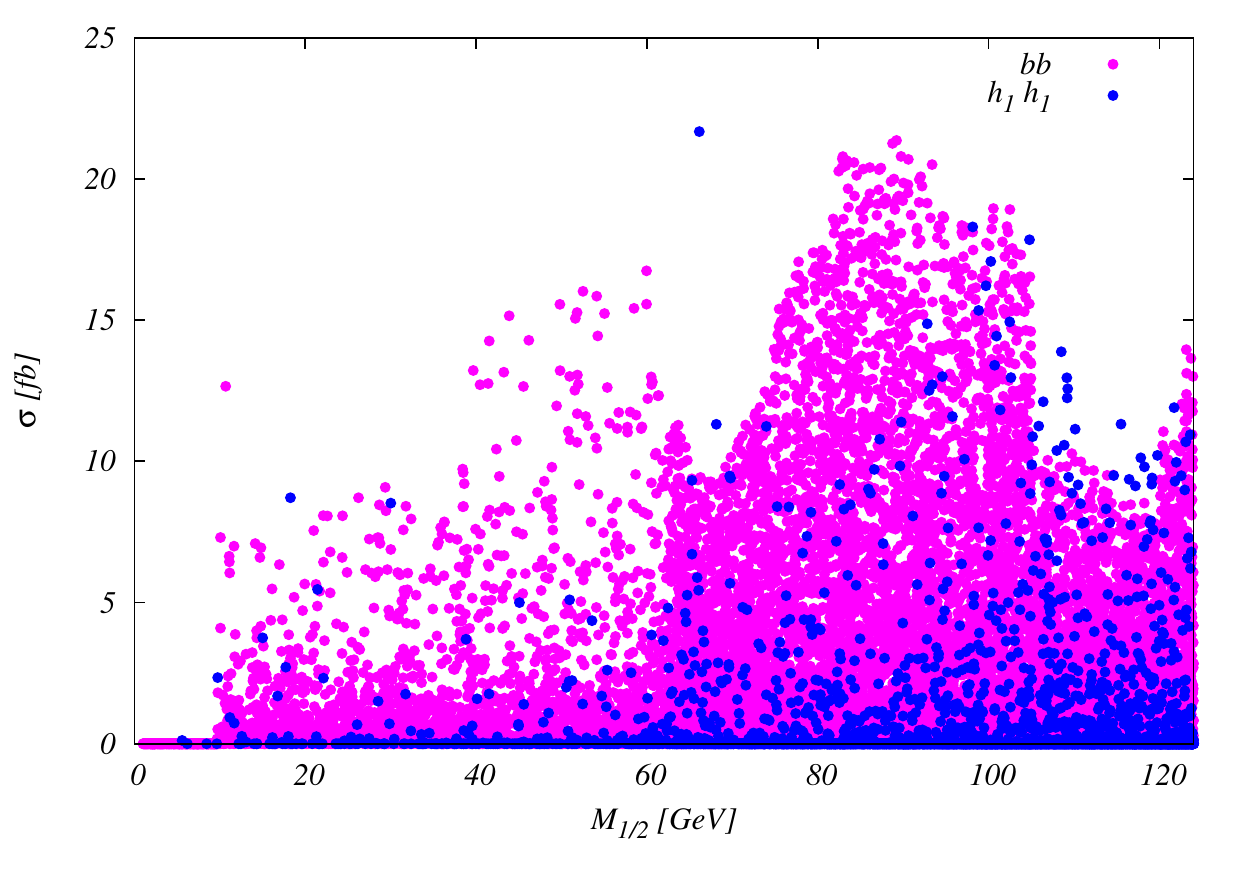}
\caption{\label{fig:bbfin} Production cross sections for $e^+e^-\,\rightarrow\,Z\,h_{1/2}\,\rightarrow\,Z\,X\,X$, with $X\,\equiv\,b$ {\sl (magenta)} and $h_1$ {\sl (blue)}. Points from all data sets are included. Cross sections can reach up to 20 \fb. In the low mass region, also $X\,\equiv\,\tau,c$ final states can become important (not shown here).}
\end{center}
\end{figure}
\end{center}

As branching ratios for the low mass scalars are inherited via mixing with the scalar from the SM-like doublet, the largest production cross sections are obtained for scenarios where the light scalars decay into $b\bar{b}$ final states. For such final states, several studies already exist projecting bounds at Higgs factories, see e.g. discussion and references in \cite{Robens:2022zgk}. We display cross sections for such final states in figure \ref{fig:bbfin}, together with predictions for $h_1\,h_1$ final states in case $h_2\,\rightarrow\,h_1\,h_1$. In the mass range $M_i\,\lesssim\,12\,\GeV$, $\tau\,\tau$ and $c\bar{c}$ final states can additionally lead to cross sections up to $20\,\fb$.

\begin{center}
\begin{figure}
\begin{center}
\includegraphics[width=0.5\textwidth]{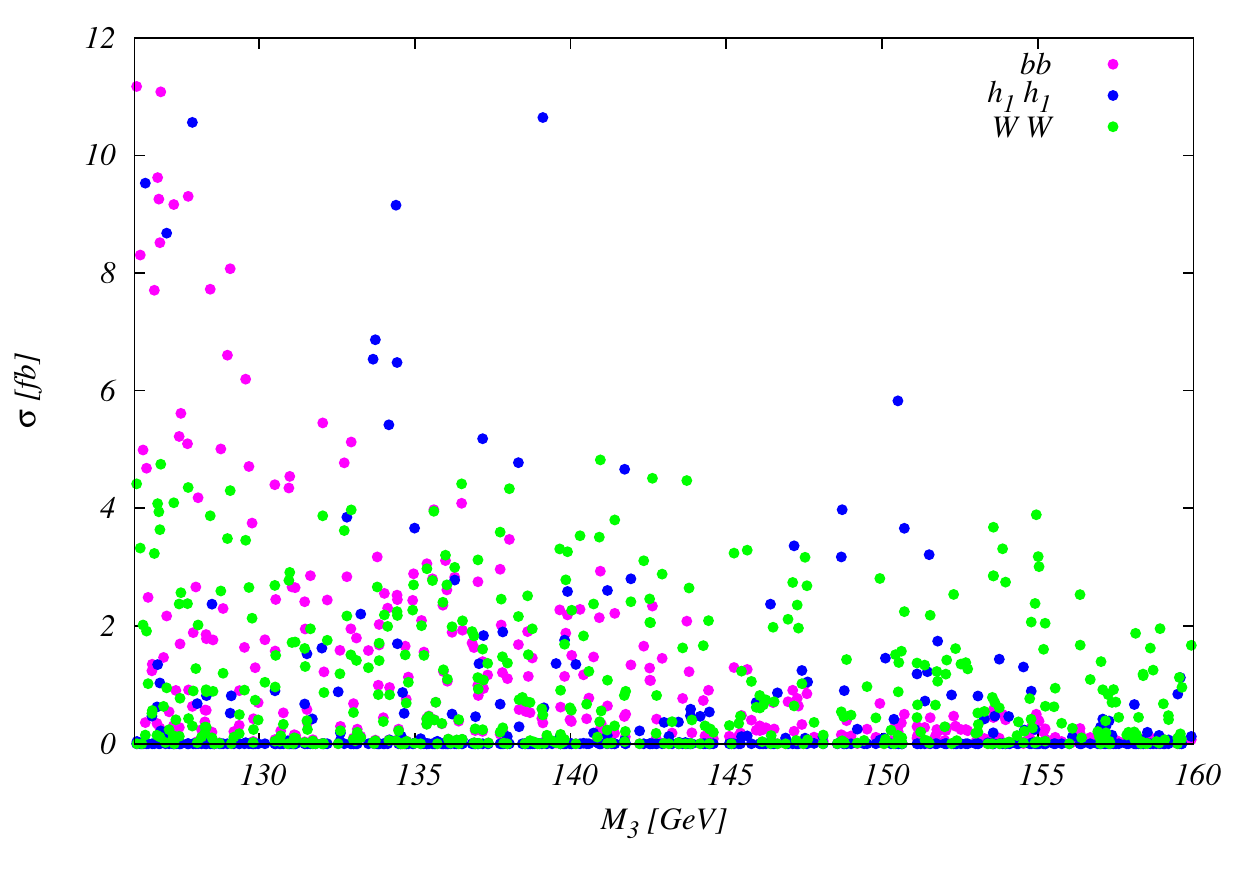}
\caption{\label{fig:finsh} Production cross sections for $e^+e^-\,\rightarrow\,Z\,h_{3}\,\rightarrow\,Z\,X\,X$, with $X\,\equiv\,b$ {\sl (magenta)}, $h_1$ {\sl (blue)}, and $W$ {\sl (green)}. Points from all data sets are included. Cross sections can reach up to $\sim\,12\,\fb$. }
\end{center}
\end{figure}
\end{center}

For the region $M_3\,\gtrsim\,126\,\GeV$, three different decay channels are dominant: $h_1\,h_1,\,W^+\,W^-,$ and $b\bar{b}$. We display the corresponding production cross sections in figure \ref{fig:finsh}. 

\begin{center}
\begin{figure}
\begin{center}
\includegraphics[width=0.5\textwidth]{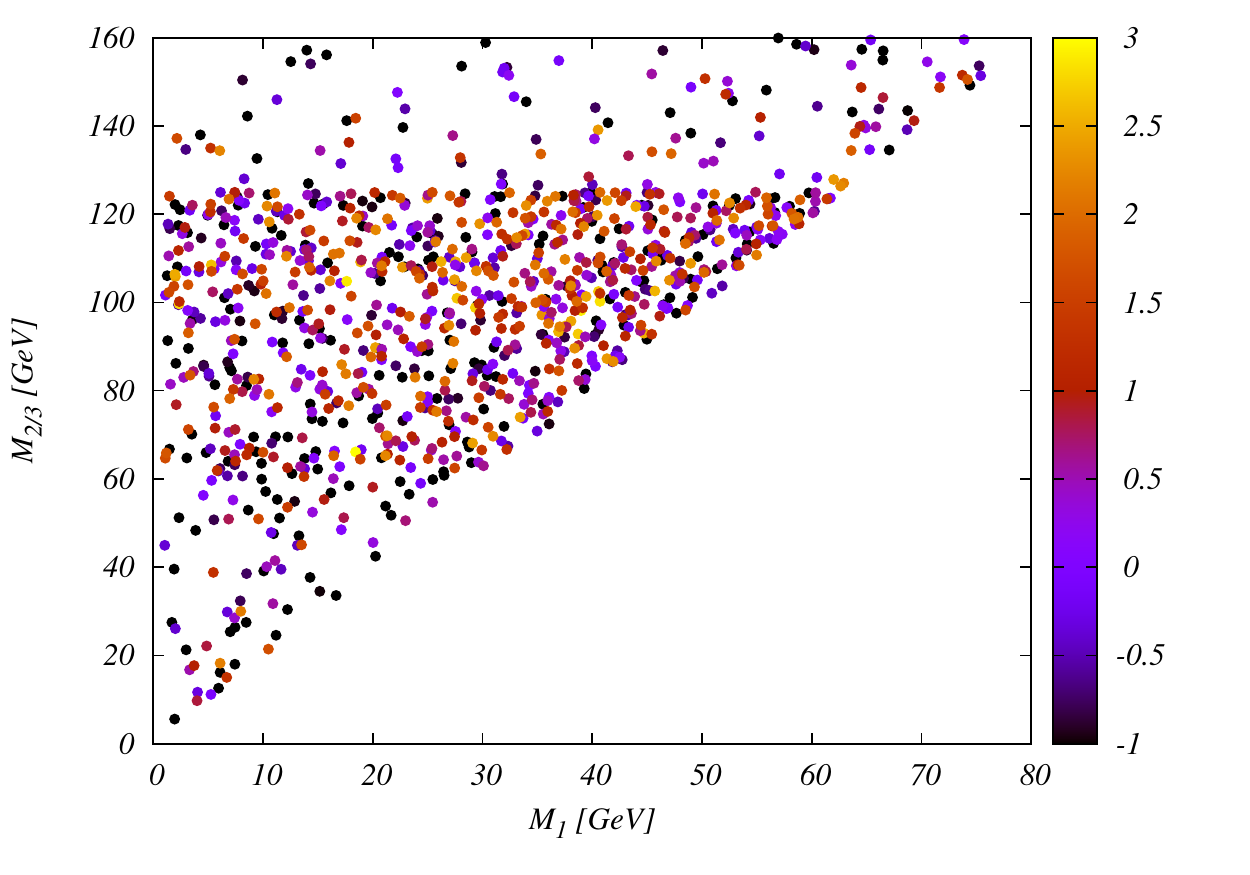}
\caption{\label{fig:Hhh} Production cross sections for $e^+e^-\,\rightarrow\,Z\,h_{2/3}\,\rightarrow\,Z\,X\,X$, with $X\,\equiv\,h_1$, in the $\lb M_1,\,M_{2/3} \rb$ plane. Color coding refers to the $\log_{10}\left[\sigma/\fb\right]$ for $Z h_1 h_1$ production. Maximal cross sections are around 20 \fb. }
\end{center}
\end{figure}
\end{center}

Finally, we can ask what cross sections can be obtained for $e^+e^-\,\rightarrow\,Z\, h_{2/3}$, with subsequent decays to $h_1\,h_1$ final states. Again ignoring cases where $M_i\,\sim\,125\,\GeV$, we display the corresponding cross sections in figure \ref{fig:Hhh}. We find the largest cross section of about $\sim\,20\,\fb$ for a parameter point where $M_2\,\sim\,66\,\GeV,\,M_1\,\sim\,18\,\GeV$. The $h_1$ in this parameter point decays predominantly into $b\,\bar{b}$ final states with a branching ratio of about $85\%$.

\section{Summary}
In this review, I gave a short summary of the status of collider signatures and searches in the TRSM introduced in \cite{Robens:2019kga}. I gave a summary on current state of the art and investigation, including further detailed collider studies, recasts, as well as current searches that use or are motivated in this model. I also gave a brief overview on channels within this model that might be testable at future $e^+e^-$ machines, with a focus in Higgs factories with $\sqrt{s}\,\sim\,250\,\GeV$. Finally, I commented on regions that would be allowed or excluded by the current value of the W-boson mass.

\bibliography{lit}

\end{document}